\documentclass[3p,times,12pt]{elsarticle}

\usepackage{amsmath}
\usepackage{amsfonts} 
\usepackage{here}
\usepackage{caption}
\usepackage{subcaption}
\usepackage{color}
\usepackage{ bbold }
\usepackage{xcolor}

\usepackage{graphicx}
\usepackage{soul}   
\newcommand{\noop}[1]{}

\usepackage{lineno,hyperref} 
\modulolinenumbers[5]
\journal{Measurement}
\usepackage{multirow}
\usepackage{ gensymb }
\usepackage{blindtext}
\usepackage{float}
\usepackage{pict2e}
\date{}

\usepackage[linesnumbered,ruled,vlined]{algorithm2e}
\usepackage{ulem}
\biboptions{sort&compress}
\usepackage[T1]{fontenc}

\usepackage{siunitx}
\usepackage{textcomp}

\newcommand{\beq}{\begin{equation}}
\newcommand{\eeq}{\end{equation}}

\usepackage{graphicx}
\usepackage{lineno,hyperref}
\modulolinenumbers[5]

\newcommand{\bx}{\mbox{\boldmath $x$}}

\newcommand{\bX}{\mbox{\boldmath $X$}}

\newcommand{\Real}{\mathbb R}

\newcommand{\be}{\begin{eqnarray}}
\newcommand{\ee}{\end{eqnarray}}

\begin{document}

\begin{frontmatter}

\title{Non-negative tensor factorization-based dependence map analysis for local damage detection in presence of non-Gaussian noise}

 \author[label1]{Anna Michalak }
 \author[label1]{Justyna Hebda-Sobkowicz}
 \author[label2]{ {Anil Kumar} \corref{cor1} }
 \author[label1]{Radoslaw Zimroz}
 \author[label3]{Rafal Zdunek}
 \author[label4]{Agnieszka Wylomanska}

 \cortext[cor1]{Corresponding author, 20210129@wzu.edu.cn}
 \address[label1]{Faculty of Geoengineering, Mining and Geology, Wroclaw University of Science and Technology, Na Grobli 15, 50-421 Wroclaw, Poland
 \\\{anna.michalak, justyna.hebda-sobkowicz, radoslaw.zimroz\}@pwr.edu.pl\\}
 \address[label2]{ {College of Mechanical and Electrical Engineering, Wenzhou University, Wenzhou, 325 035, China
 \\ 20210129@wzu.edu.cn}\\}
 \address[label3]{Faculty of Electronics, Photonics, and Microsystems, Wroclaw University
 of Science and Technology, Wroclaw, Poland\\rafal.zdunek@pwr.edu.pl}

 \address[label4]{Faculty of Pure and Applied Mathematics, Hugo Steinhaus Centre, Wroc{\l}aw University of Science and Technology, Hoene-Wronskiego 13 C, 50-376 Wroclaw, Poland\\agnieszka.wylomanska@pwr.edu.pl}

\begin{abstract}

 {The time-frequency map (TFM) is frequently used in condition monitoring, necessitating further processing to select an informative frequency band (IFB) or directly detect damage. However, selecting an IFB is challenging due to the complexity of spectral structures, non-Gaussian disturbances, and overlapping fault signatures in vibration signals. Additionally, dynamic operating conditions and low signal-to-noise ratio further complicate the identification of relevant features that indicate damage. To solve this problem,  the present work proposes a novel method for informative band selection and local damage detection in rolling element bearings, utilizing non-negative tensor factorization (NTF)-based dependence map analysis. The recently introduced concept of the dependence map is leveraged, with a set of these maps being factorized to separate informative components from non-informative ones. Dependence maps provide valuable information on the auto-similarity of spectral content, while NTF, a powerful tool commonly used in image processing for feature extraction, enhances this process. The combination of these methods allows for the extraction of IFBs, forming the basis for local damage detection. The effectiveness of the proposed method has been validated using both synthetic and real vibration signals corrupted with non-Gaussian disturbances.}

\end{abstract}

\begin{keyword}
fault detection, bearings, vibration, measures of dependence, non-negative tensor factorization
\end{keyword}

\end{frontmatter}

\section{Introduction}
The detection of local damage in rotating machinery is an important topic \cite{randall2011rolling}. {Local damage in the context of rolling element bearings or gears refers to defects occurring in specific regions {such as the races or rolling elements} of the bearing rather than damage distributed over a wider area. {Although industrial systems are often very complex, bearing failures usually start with localized damage, and the analysis aims to identify such early defects in the context of complex operating conditions.} Due to the rotational character of the object of interest, it produces a cyclic and impulsive signal of interest (SOI).} Unfortunately, the SOI is often masked by other sources in complex mechanical systems. A key issue is to detect a weak SOI (associated with a fault at an early stage of development) in noisy observation (vibration, acoustic signal) measured from the machine for specific conditions, namely, in the presence of non-Gaussian, impulsive noise. As the problem of local damage detection is important, many work has already been published. Pre-filtering of the raw signal may be identified as the main direction of research. The pre-filtered signal has a better signal-to-noise ratio (SNR), and it is easier to detect damage.

The most popular methods for band selection are spectral kurtosis and kurtogram \cite {antoni2006spectral, antoni2007fast}. Both approaches exploit kurtosis as a measure of impulsiveness. In the spectral kurtosis approach, the signal is spectrally decomposed into a family of subbands via a spectrogram, next for each band the sample kurtosis is estimated, and finally, this statistic as a function of frequency is used to select an impulsive contribution. In a fast kurtogram, the decomposition is based on the filterbank approach. Both approaches are continuously being improved by various authors \cite{wang2013enhanced,peter2013design,wang2016new,liu2019accugram,moshrefzadeh2018autogram,mauricio2020improved,wu2021enkurgram}.
Many possible statistics have appeared to be used for band selection, i.e., parameter $\alpha$ from $\alpha$-stable distribution \cite{Yu2013155}, spectral Gini index \cite{miao2017improvement} or conditional variance \cite{Hebda-Sobkowicz2020mssp}. The filter characteristic could also be estimated using optimization techniques. For example, Wodecki proposed a genetic algorithm with kurtosis as a cost function for filter design \cite{wodecki2018optimal}.
The signals from rotating machine faults are both impulsive and cyclic. To take advantage of both features, Antoni proposed a method for fault detection called infogram \cite{antoni2016info}. It has also been extended by Wang \cite{wang2016extension} and Hebda-Sobkowicz \cite{hebda2022infogram} to be more robust in case of non-Gaussian noise in the signal. {In \cite{cheng2022improved}, Cheng et al. proposed the improved envelope spectrum via candidate fault frequencies optimization-gram (IESCFFOgram) to identify the IFB using spectral coherence.}

 {However, used measures of impulsiveness may fail to extract the signal of interest if the information is hidden in non-Gaussian heavy-tailed distributed (impulsive) noise and SNR is poor. In this situation, the measures respond to the bands occupied by the largest impulses.} Thus, several techniques have been developed to deal with \cite{Yu2013155,borghesani2017cs2,hebda2022infogram, Mauricio2020, HebdaSobkowicz2020, kruczek2020detect,zhou2023cffsgram,cui2024spectral}. 
They are based on robust statistics for band selection, generalization of cyclostationary analysis by replacement estimators or measures of dependence, logarithmic stabilization of envelope, etc.
A recent review of existing methods can be found in \cite{HebdaSobkowicz2020,yang2024review}.

As mentioned, the time-frequency representation (TFR) is often used to visualize and analyze the spectral content of the acquired signal \cite{allen1977short}. The spectrogram can be processed in many ways \cite{FabienM,4480140}. The mentioned selectors allow for the identification of specific properties in some frequency bands. However, it is also possible to detect some events in a time-frequency map or even extract them. As the time-frequency map is a non-negative matrix describing the fluctuation of energy in each frequency band in time, recently, Wodecki \cite{wodecki2019impulsive,Wodecki2020,wodecki2021local} and Gabor \cite{Gabor20242944} proposed to use the processing of the spectrogram by non-negative matrix factorization (NMF). It allows both to identify IFB and the location of the events in time.
NMF introduced by Lee and Seung \cite{lee1999learning}, is a powerful subspace decomposition method for dimensionality reduction and feature extraction. It has been developed by many research groups for various applications, including techniques for processing vibration signals, achieved by iteratively performing matrix factorization to extract key data features.
 Other applications of NMF can be found in \cite{wang2015non,liang2016feature,wang2021intelligent}. One of the extensions of NMF is non-negative tensor factorization (NTF) \cite{Shashua2005,carroll1989fitting} that is addressed for multilinear feature extraction from non-negative data represented in the form of a multi-way array. This model has already found multiple applications in signal processing; see, e.g., \cite{FitzGerald2008,6588559,8497054,app9183642}. 
\textcolor{black}{In \cite{Gabor2023} was shown that the NTF model significantly outperforms NMF for noisy synthetic and real data. }

{With the development of methods based on time-frequency representations, different variants of them are also being proposed, constantly improving signal analysis capabilities. This is especially significant for non-stationary conditions. In \cite{Zhao2024335}, Zhao et al. described a time-frequency analysis (TFA) technique called CTNet, which was designed to effectively characterize non-stationary signals with closely spaced or crossing instantaneous frequencies (IFs), particularly in the context of fault diagnosis in wind turbines. CTNet is capable to accurately amplify the key features in a TFR. 
In addition, in \cite{ZHAO2024111112}, a technique called frequency-chirprate synchrosqueezing-based scaling chirplet transform (FCSSCT) was proposed. It introduced the chirprate dimension, creating a three-dimensional time-frequency-chirprate space that allowed a precise representation of non-stationary signals and better detection of closely spaced or crossing frequencies.}

Cyclostationary analysis is one of the most intuitive methods for rotating machines, as damaged bearings or gears produce a cyclic impulsive signal \cite{randall2011rolling}.
This bi-frequency cyclostationary representation has also been processed using NMF \cite{wodecki2019impulsive}.
When vibration or acoustic signal captured from the faulty component contains some non-Gaussian, impulsive disturbance, the classical cyclic spectral coherence (CSC) map may not be as efficient as expected.{ To investigate alternative measures of dependence, Nowicki et al. \cite{nowicki2021dependency} proposed several methods specifically designed for analyzing the time-frequency map.} As a result, a novel bi-frequency map can be obtained that contains similarity measures between subsignals for each pair of bands $\Delta f_i-\Delta  f_j$. For noisy bands, a similarity tends to zero, for bands with similar pattern, it will converge to one.
Unfortunately, for non-Gaussian noise, it may also highlight some bands with impulsive disturbances with high correlation, so there is a need to separate information related to various sources.
 
{Traditional methods like spectral kurtosis struggle with non-Gaussian, impulsive noise, confusing it with faults  \cite{antoni2006spectral, antoni2007fast}. In contrast, the proposed Non-negative Tensor Factorization (NTF) with $\beta$-divergence effectively separates the signal of interest (SOI) from impulsive disturbances \cite{Gabor2023}. In band selection, traditional methods treat frequency bands independently \cite{antoni2006spectral,Yu2013155,miao2017improvement,Hebda-Sobkowicz2020mssp} while the proposed approach constructs a dependency map to improve identification of informative bands \cite{nowicki2021dependency}. Classical methods, like CSC, in the case of impulsive non-Gaussian noise, experience a significant decrease in quality, which makes it impossible to detect the periodic signal \cite{wodecki2021influence}. Crucially, the proposed approach operates unsupervised, requiring no prior knowledge of faults, making it adaptable to various machinery and conditions. It effectively addresses challenges in non-stationary and noisy industrial environments, particularly in applications like crushers and rolling element bearings.}

{In recent years, bearing diagnostics have seen a significant shift toward data-driven approaches, with machine learning techniques such as neural networks, support vector machines (SVM), support matrix machines (SMM), and others \cite{zheng2024progressive, ye2023intelligent, wissbrock2024more, Han2021, Li20227328, Luo2024}. In \cite{zheng2024progressive}, Zheng et al. proposed a progressive multisource adaptation method based on matching domain-independent and category-informative features to improve fault diagnosis performance in cross-domain scenarios.
In \cite{Han2021}, Han et al. combined a convolutional neural network (CNN) with a support vector machine (SVM) to improve bearing fault diagnosis with a small number of samples, while in \cite{Li20227328} used least square interactive support matrix machine (LSISMM) matrix classifier to analyze thermal images under variable speed conditions.
Although these methods face challenges such as the need for large data sets and significant computational resources, they provide many advantages that encourage their use.}

Diagnosing bearings in a copper ore crusher is particularly difficult due to the non-stationary and non-Gaussian nature of signals. These challenges arise from variable loads caused by fluctuating ore characteristics, which create inconsistent operational conditions. Additionally, high-impact forces during crushing introduce abrupt changes, while non-Gaussian noise - characterized by heavy tails and impulsive spikes - makes it difficult to distinguish between noise and actual faults.

In this paper, the NTF is applied to such a bi-frequency map for IFB signature extraction. {The proposed technique allows the identification of critical frequency bands essential for damage detection amidst overlapping fault signatures and non-Gaussian disturbances. By isolating informative from non-informative components, the NTF approach increases robustness against the noise typical in the crusher environment. Furthermore, the use of short-time Fourier transform (STFT) allows dynamic adaptation to variable operational loads.} To extract the NTF factors, the $\beta$-divergence was used, which is minimized with the multiplicative update rules \cite{cichocki2009nonnegative}. The choice of this objective function is motivated by its greater flexibility in adapting to the separation problem with heavy-tailed non-Gaussian residual errors \cite{Gabor2023}. 
In general, the signal has been segmented
along the time axis into $M=30$ blocks of $P$ samples. For each segment, the spectrogram and then the Pearson dependence map were calculated. Then, the 3-way array (frequency $\times$ frequency $\times$ dependency map) is processed with the NTF algorithm with various $\beta$ parameters. 
 This multilinear approach allows us to tailor the blind source separation tool to extract SOI from a mixed signal registered from the bearings of the rolling element. It was assumed that the periodic and impulsive SOI is perturbed with a non-periodic sparse impulse noise or some non-stationary component, and a strong independent and identically distributed Gaussian noise. The dependency map will highlight information about similarities between various frequency bands, indicating in this way bands with high and poor correlation. {The use of maps from subsequent segments makes it possible to reduce the impact of non-cyclic pulses (since they vary over time) and to obtain robust results. By applying NTF to the set of dependency maps, the proposed approach allows to obtain a selector that correctly indicates the informative frequency band. The proposed procedure produces high-quality results and allows the detection of local damage at an early stage when most known methods have failed.}

The paper is organized as follows. First, the NTF-based dependence map analysis framework for local damage detection is defined, see Section \ref{method}.  {In this Section, definitions of all essential components are reviewed.}
Next, Section \ref{simul} describes the numerical experiments to obtain synthetic signals (simulations) and the primary results of our method for simulation signals.
The experimental setup for the acquisition of vibration signals and the result of the proposed method for real signals are provided in Section \ref{real_analysis}.
 {The last subsections, Section \ref{com_sim} for simulated data and Section \ref{com_real} for real data, include comparisons with several recently developed frequency band selectors. These selectors encompass both classical methods and more robust approaches based on spectral kurtosis, the parameter $\alpha$ from $\alpha$-stable distribution, conditional variance, and Pearson correlation. Spectral kurtosis was chosen due to its classical popularity, while the next two were selected for their effectiveness with data contaminated by non-Gaussian noise. Additionally, a selector based on Pearson’s correlation coefficient was proposed, as it motivated the proposed procedure and the tensor's dependency map construction relies on it. A discussion emphasizing the advantages of NTF is presented here, with conclusions provided in Section \ref{concl}.}

\section{Methodology} \label{method}

 {The general concept of the work is to detect local damage of the rolling element bearings using the vibration signal. The challenge of the detection problem is related to a weak cyclic impulsive signal (early stage of the damage) and strong, non-Gaussian (impulsive) noise which often fully covers the signal of interest. Both components are non-stationary; thus, using time-frequency analysis is justified. However, the small SNR makes it difficult to detect fault-related patterns in the spectrogram. It should be highlighted that the location and amplitude of non-cyclic impulses are completely unknown and random. {They are related to the hard operating conditions of the machine and cause the vibration signal obtained to be strongly non-stationary. } It is important to note that the method is designed to handle non-Gaussian noise, and this makes it unique compared to many classical methods. This adaptability ensures the applicability of the method in a wide range of real-world scenarios. {In summary, the signal under consideration is strongly non-stationary due to its complex design and the nature of the machine's operation (falling ore fragments in the case of a crusher). The purpose is to create a procedure that, under these difficult conditions, can detect the damage at an early stage and plan the repair without the risk of an emergency breakdown of production.} The whole procedure is unsupervised and does not require training or assumptions. As a result of tensor decomposition, one may obtain a filter characteristic that matches the informative frequencies.}

 {To address this problem, the presented methodology proposes using the concept of a dependence map. It assumes that the signal is decomposed spectrally (via a spectrogram) into a set of narrowband time series, and a similarity estimator is calculated for each pair of signals. If the signals are different (not similar), the coefficient will tend to be zero. If the variation of the amplitudes of two signals is somehow synchronized, the similarity will be close to one.
 
To provide a statistical background of the method, it is proposed to divide the signal into multiple segments and to transform each segment into a time-frequency representation.  Next, similarities between spectral bands have been aimed to be found using dependence measures, and a novel bi-frequency representation has been created through a dependence map. It could be seen as a kind of data fusion applied in order to extract information about the repeating pattern. Finally, the developed data structure, consisting of multiple bi-frequency representations, has been decomposed using tensor analysis.

The NTF method allows us to decompose the 3D array structure into 3 matrices. As shown in \cite{Gabor2023}, one of them will enable us to identify the frequency band. In the case of a bi-frequency symmetric map (not time-frequency as in \cite{Gabor2023}) two of them provide the same information. }

\begin{figure}[h!]
    \centering
    \includegraphics[width=\textwidth]{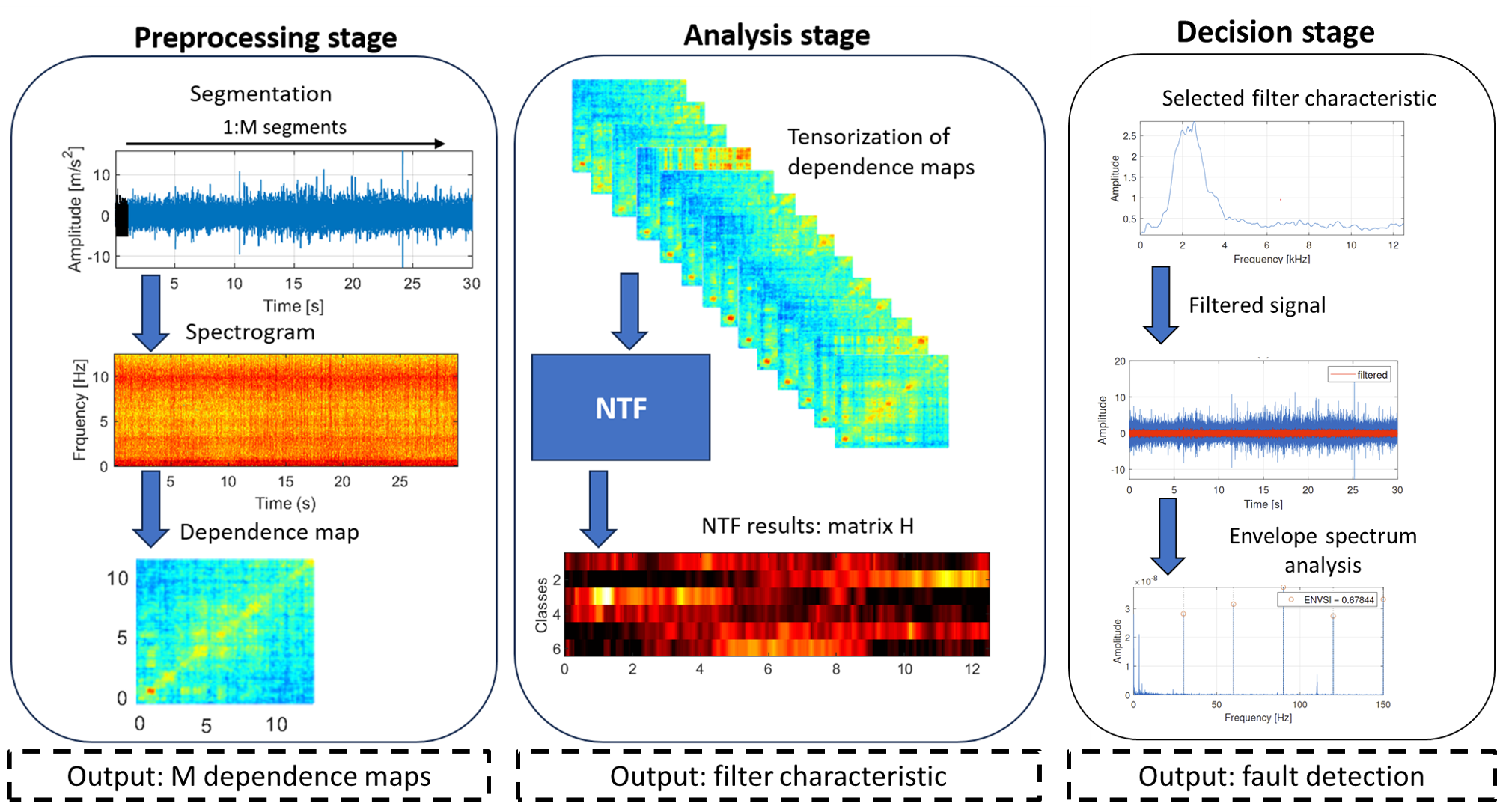} \\
    \caption{Flowchart of the proposed method}
    \label{fig:flowcharts}
\end{figure}

 {In Fig. \ref{fig:flowcharts}, the procedure is presented in the form of a diagram. {It consists of {three} main stages: pre-processing, analysis, and the decision.}}

 {  \begin{itemize}
    \item \textbf{Pre-processing stage} is to prepare a specific data structure consisting of multiple bi-frequency maps. To achieve this, the signal is divided into M segments. Then, for each segment, the spectrogram is calculated. It might be replaced with more advanced time-frequency representations with better resolutions. The spectrogram is then considered as a set of time series related to each frequency band, and by searching for relationships between bands (for each pair), a dependence map is created to describe the similarity between two given bands from the spectrogram. Finally, a specific data structure called a tensor is obtained, which consists of bi-frequency representations of M segments, as shown in Fig. \ref{fig:flowcharts}.
    \item During \textbf{analysis stage}, the tensor decomposition is applied, and the data structure is decomposed by using NTF into three matrices. One of these matrices will be further used for identifying informative bands. The matrix H presented in Fig. \ref{fig:flowcharts} shows several classes of information distribution (Y axis) along the frequency axis (X axis). 
    \item {In the \textbf{decision stage}, one of these classes is chosen as a filter characteristic to further filter a raw signal and obtain a Signal of Interest. This selection is made by using the maximum value of the Envelope Spectrum Based Indicator (ENVSI) of the filtered signals. Finally, the decision regarding the damage of the component in question is made by assessing the value of the ENVSI. The ENVSI ranges between 0 and 1, with higher values indicating a greater energy in the component corresponding to the damage, and values close to zero suggesting the absence of damage. Based on the analysis of the real vibration signal, one may conclude that an ENVSI value greater than 0.1 may lead to information about the occurrence of the SOI. }
\end{itemize}
}

Notation: Multi-way arrays, matrices, vectors, and scalars are denoted by uppercase calligraphic letters (e.g., $\mathcal{X}$), uppercase boldface letters (e.g., $\bX$), lowercase boldface letters (e.g., $\bx$), and unbolded letters (e.g., $x$), respectively. Following the notation of Kolda \cite{Kolda08}, multi-way arrays will be referred to as tensors.

\subsection{Spectrogram}

The time-frequency representation of the signal is widely used since it gives the temporal and spectral content simultaneously and provides much more information about the data than the one-dimensional representation in the time domain. Signal decomposition can be performed by the short-time Fourier transform  or many other techniques (Wavelets, Wigner, EMD, etc.); see, e.g.~\cite{feng2013recent}. In this paper, STFT is used.

Let $\textbf{x}=(x_1,\dots,x_N)$ be the input vibration signal of length $N$, which can be partitioned into $M$ blocks of length $P$, where $N = MP$. Following this procedure, $\textbf{X}=[\textbf{x}_1,\dots,\textbf{x}_M]\in \Real^{P \times M}$.  Each segment of the signal $\mathbf{x}$, i.e., $[\textbf{x}_1,\dots,\textbf{x}_M]$, is transformed into the time-frequency domain by STFT according to the following definition \cite{allen1977short}:

\begin{equation}
    STFT_x(f,t)=\sum_{h=1}^{P}x_h \gamma _{t-h}e^{\frac{-2j\pi fh}{P}},
    \label{eq:stft}
\end{equation}
where $\gamma _{t-h}$ is the shift window, $\textbf{x}_m=(x_{1},\dots,x_{P})$ is the $m^{th}$ segment of the input signal of length $P$, $t=t_1,\dots,t_T$ is the time point, $f=f_1,\dots,f_F$ is a frequency bin, and $j$ is an imaginary unit; see \cite{boash} for more details. The spectrogram is defined as the absolute values of STFT, that is, $\mathbf{S}(f,t)=|STFT(f,t)|\in \Real^{T\times F}$.
{During data analysis, the commonly used Hamming window of the length of 256 samples and the overlap equal 85\% was applied, and the Fourier transform was calculated for 512 frequency points.}

\subsection{Pearson correlation map construction}
\label{sec:pea}

The most well-known measure of dependence is the
Pearson correlation coefficient. Let consider the spectrogram $\mathbf{S}=[\mathbf{s}_{f_1},\dots,\mathbf{s}_{f_F}] $ as a set of vectors, then the Pearson correlation for the vectors $\mathbf{s}_{f_j}$ and $\mathbf{s}_{f_k}$ is defined as follows \cite{dunn2009basic}: 

\begin{equation}
\hat{\rho}(\mathbf{s}_{f_j},\mathbf{s}_{f_k}) =\frac{ \sum_{ i=1 }^{ T } (\mathbf{s}_{f_ji} - \overline{\mathbf{s}}_{f_j})(\mathbf{s}_{f_ki} - \overline{\mathbf{s}}_{f_j}) }{ \sqrt{ \sum_{ i=1 }^{ T } (\mathbf{s}_{f_ji} - \overline{\mathbf{s}}_{f_j})^2} \sqrt{ \sum_{ i=1 }^{ T } (\mathbf{s}_{f_ki} - \overline{\mathbf{s}}_{f_k})^2} },
\label{pearson}
\end{equation}
where $\overline{\mathbf{s}_{f_j}} $, $ \overline{\mathbf{s}_{f_k}} $ are sample means of data vectors $\mathbf{s}_{f_j}$ and $\mathbf{s}_{f_k}$, respectively.

The Pearson correlation is used to investigate the relationship between two variables. It is worth mentioning that only the linear relationship is investigated. The Pearson correlation coefficient values are in the range $[-1,1]$. Values close to $1$ and $-1$ indicate a strong relationship (positive and negative, respectively). A value of $0$ informs about no dependency. The Pearson correlation coefficient is known to be sensitive to outliers. Therefore, one can observe its high values for frequency bands with high-energy non-cyclic impulses.

The slices of the spectrogram, i.e. $\mathbf{S}$ that correspond to the frequency bins $f_i$, are used to test the correlation and the symmetric correlation map (\textbf{CM}) is created, i.e.:
 \begin{equation}
\label{eq:sab}
\textbf{CM}(j,k)=\hat{\rho}(\mathbf{s}_{f_j},\mathbf{s}_{f_k}), \text{ 
 for } j,k=1,\dots,F.
 \end{equation}
 The function $\hat{\rho}(\mathbf{s}_{f_j},\mathbf{s}_{f_k})$ denotes the empirical Pearson correlation as described in Eq. (\ref{pearson}). 
 If the slices $\mathbf{s}_{f_j}$ and $\mathbf{s}_{f_k}$ reveal a similar behavior, then they should be treated as the same source of information and the correlation value tends to 1. As $\mathbf{CM}$ is symmetric, only half of the matrix value is needed for the calculation.

To investigate the existence of correlation in the time-frequency representation of the signal, the calculation of the correlation is performed by different correlation measures because some measures are sensitive to outliers \cite{nowicki2021dependency}.
As presented in \cite{nowicki2021dependency}, there are many possible dependence measures that could be used here. 

However, the most popular Pearson correlation is used and takes advantage of the tensorization of many Pearson maps.

\subsection{Non-negative Tensor Factorization}

Let $\textbf{x}=(x_1,\dots,x_N)$ be the input vibration signal of length N, which can be divided into M segments. For each of them, a spectrogram ($\textbf{S}\in \Real_+^{T \times F}$) and then a correlation map ($\textbf{CM}\in \Real^{F \times F}$) are calculated, respectively. Due to the fact that Pearson correlation can take negative values, and NTF operates only on non-negative arrays, the absolute value of $\textbf{CM}$ is used i.e. $\textbf{C}=|\textbf{CM}|\in \Real_{+}^{F \times F}$. Subsequent \textbf{C} matrices are arranged into a third-order tensor $\mathcal{C}=[\textbf{C}_1,\dots,\textbf{C}_M] \in \Real_+^{F \times F \times M}$, where the first two dimensions represent frequencies, and the third dimension corresponds to M segments since the correlation map is symmetric and defined as the correlation for every two frequency bins.

 \begin{figure}[h!]
     \centering
     \includegraphics[scale=0.35]{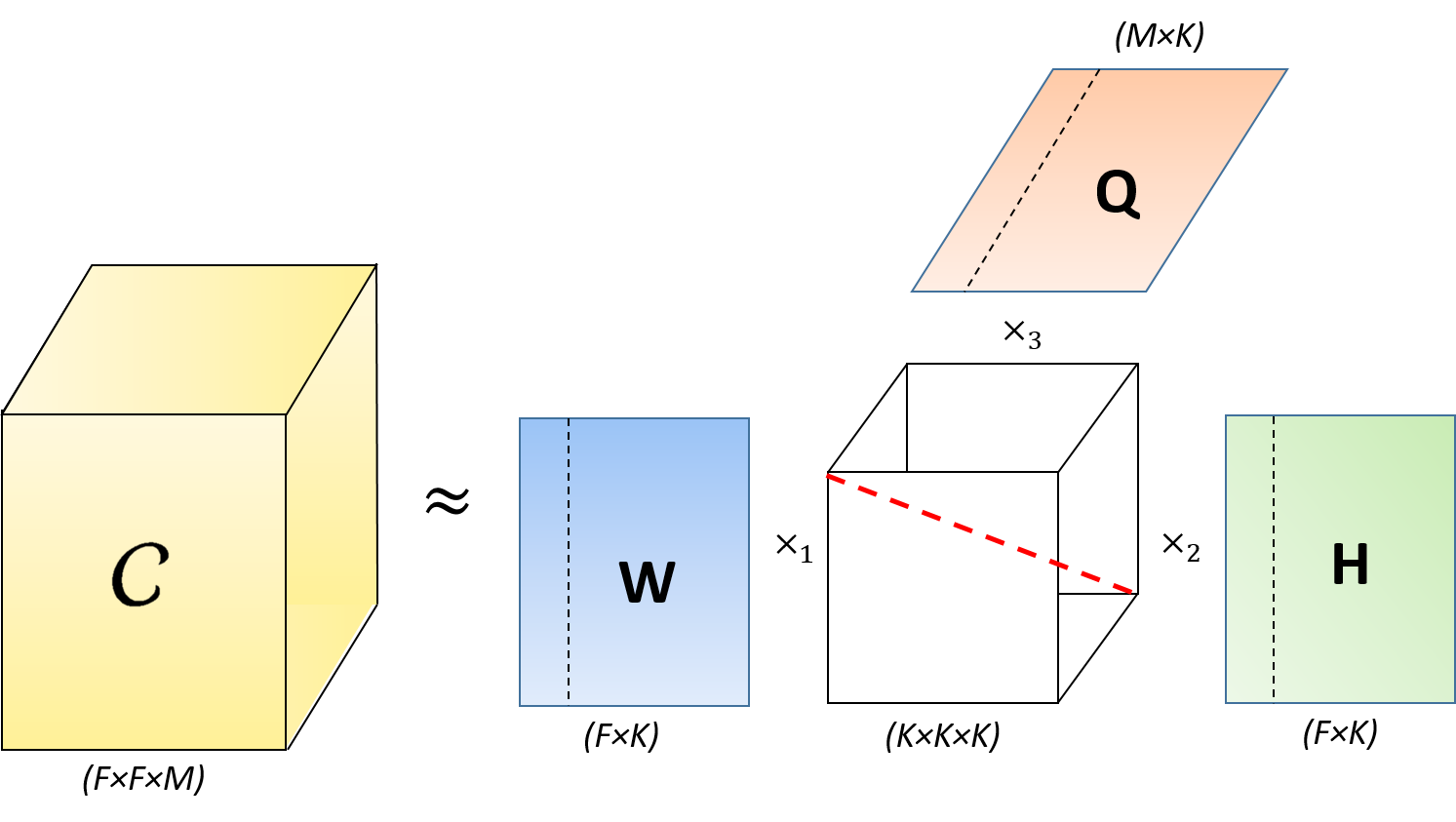}
     \caption{NTF factorization of the 3-order tensor $\mathcal{C}$.}
     \label{fig:NTF}
 \end{figure}
 
In general, the NTF model for the three-order tensor $\mathcal{C} \in \Real_+^{F \times F \times M}$ can be defined as follows \cite{Kolda08,cichocki2009nonnegative}:

\begin{equation}
    \mathcal{C} = \sum_{k=1}^{K} (\textbf{w}_{k} \circ  \textbf{h}_{k} \circ \textbf{q}_{k})+ \mathcal{E} = \mathcal{I} \times_1 \textbf{W} \times_2 \textbf{H} \times_3 \textbf{Q} + \mathcal{E},
\end{equation}
where $\textbf{W}=[\textbf{w}_1,\dots,\textbf{w}_K] \in \Real_+^{F \times K}$, $\textbf{H}=[\textbf{h}_1,\dots,\textbf{h}_K] \in \Real_+^{F \times K}$, and $\textbf{Q}=[\textbf{q}_1,\dots,\textbf{q}_K] \in \Real_+^{M \times K}$ are three non-negative component matrices, $\mathcal{E}= \mathcal{C} - \hat{\mathcal{C}}$ is a tensor representing the error, the upper hat $\hat{\mathcal{C}}$ is the estimator of $\mathcal{C}$, $K$ is the NTF rank, and $\circ$ is the outer product, and $\times_n$ is the n-mode product.

\subsubsection{Algorithmic approach}

Many of the well-known NTF algorithms utilize alternating minimization of the squared Euclidean distance or the generalized Kullback-Leibler divergence \cite{berry2007algorithms,sra2005generalized}. However, in this particular scenario, a more general cost function, known as $\beta$-divergence, is examined \cite{cichocki2009nonnegative}.
The $\beta$-divergence for the three-way NTF model can be defined as:

\begin{equation}\label{eq:beta}
   \Psi^{(beta)} = \left \{ \begin{array}{cc} 
\displaystyle{ \sum_{f_1,f_2,m} \left (c_{f_1f_2m} \frac{c_{f_1f_2m}^{\beta} - \hat{c}_{f_1f_2m}^{\beta}}{\beta} - \frac{c_{f_1f_2m}^{\beta+1} - \hat{c}_{f_1f_2m}^{\beta+1}}{\beta+1}   \right )} & \beta  > 0, \\
\displaystyle{ \sum_{f_1,f_2,m} \left (c_{f_1f_2m} \ln \left(\frac{c_{f_1f_2m}}{\hat{c}_{f_1f_2m}} \right) - c_{f_1f_2m} + \hat{c}_{f_1f_2m}   \right )} & \beta = 0, \\
\displaystyle{ \sum_{f_1,f_2,m} \left ( \ln\left(\frac{\hat{c}_{f_1f_2m}}{c_{f_1f_2m}} \right)  + \left(\frac{c_{f_1f_2m}}{\hat{c}_{f_1f_2m}} \right) - 1 \right )} & \beta = -1,
\end{array}  \right. 
\end{equation}
where $c_{f_1f_2m}$ is the single element of the tensor $\mathcal{C}$ on position $(f_1,f_2,m)$, $f_1=1,\dots,F$ and $f_2=1,\dots,F$ are the iterators across the first and second frequency dimension, respectively, $m=1,\dots,M$ is the iterator across the dimension corresponding to the number of maps and the $\hat{c}_{f_1f_2m}=\sum_{k}w_{f_1k}h_{f_2k}q_{mk}$.
For some beta parameters, the $\beta$-divergence corresponds to known distances. When $\beta = -1$, it simplifies to the Itakura-Saito (IS) distance. It is most appropriate for Gamma-distributed data. If $\beta=0$, the $\beta$-divergence takes the form of a generalized Kullback-Leibler (KL) divergence. It is optimal for data with a Poisson distribution. In the case of a Gaussian distribution, an optimal value of the $\beta$ parameter is 1. 

This cost function allows us to consider a wider class of NTF algorithms. To derive multiplicative learning rules for NTF, the gradient of the  $\beta$-divergence (Eq. (\ref{eq:beta})) with respect to $w_{f_1k}$ is computed as \cite{cichocki2009nonnegative}:
\begin{equation}
\frac{\partial \Psi^{(beta)}}{\partial w_{f_1k}}= \sum_{f_2,m}\left(\hat{c}^{\beta}_{f_1f_2m} - c_{f_1f_2m}\hat{c}^{\beta-1}_{f_1f_2m}  \right)q_{mk}h_{f_2k}
\end{equation}

The multiplicative $\beta$ NTF update rules, which are obtained according to the first-order gradient descent procedure with suitable learning rates, can be defined as follows \cite{cichocki2009nonnegative}:
\begin{equation}
w_{f_1k} \leftarrow w_{f_1k}\frac{\sum_{f_2,m}h_{f_2k}q_{mk} \left( \frac{c_{f_1f_2m}}{\hat{c}^{1-\beta}_{f_1f_2m}} \right)}{\sum_{f_2,m}\hat{c}^{\beta}_{f_1f_2m}h_{f_2k}q_{mk}}, \qquad \forall_{f_1,k},
\end{equation}

\begin{equation}
h_{f_2k} \leftarrow h_{f_2k}\frac{\sum_{f_1,m}w_{f_1k} q_{mk} \left( \frac{c_{f_1f_2m}}{\hat{c}^{1-\beta}_{f_1f_2m}} \right)}{\sum_{f_1,m}\hat{c}^{\beta}_{f_1f_2m}w_{f_1k} q_{mk}} \qquad \forall_{f_2,k},
\end{equation}

\begin{equation}
 q_{mk} \leftarrow  q_{mk}\frac{\sum_{f_1,f_2}w_{f_1k}h_{f_2k} \left( \frac{c_{f_1f_2m}}{\hat{c}^{1-\beta}_{f_1f_2m}} \right)}{\sum_{f_1,f_2}\hat{c}^{\beta}_{f_1f_2m}w_{f_1k}h_{f_2k}}. \qquad \forall_{m,k}.
\end{equation}
\label{eq_ntf_unfold_Yn}

\subsection{Envelope Spectrum Based Indicator}
To compare the efficiency of the filtering procedure, ENVSI  is used \cite{HebdaSobkowicz2020}.
In the paper, the squared envelope spectrum (SES) of the signal is used to compare the portion of the energy related to fault frequency (amplitude of informative impulses - AIS) and its harmonics 
to the whole energy of the squared envelope spectrum. The higher the index value, the better the filtering result of a tested selector.

The considered indicator is defined as \cite{HebdaSobkowicz2020}: 
\begin{equation}\label{eq:ENVSI}
    ENVSI=\frac{\sum_{i=1}^{R_1}{(AIS_i)^2}}{\sum_{k=1}^{R_2}
    (SES_k)^2},
\end{equation}
where $R_{1}$ is the number of components to analyze corresponds to the impulses of fault frequency and $R_{2}$ is the number of frequency bins used to calculate the total energy. Note that the last component of the amplitude of informative impulses ($AIS_{R_1}$) is the same component of the envelope spectrum ($ES_{R_2}$) used to calculate the indicator. The number of harmonics used for the ENVSI calculation is arbitrarily chosen (set at 5), based on the total shape of the SES. 

The lack of an impulsive component in the SES considered implies that ENVSI converges to zero. When impulses appear in the time domain, informative components are present in the squared envelope spectrum, and the ENVSI value increases. Depending on the level of the background noise, the value of ENVSI might be higher (low noise, i.e. visible impulses in the time domain)  or lower (barely visible impulses due to a high level of noise).
ENVSI makes it possible to compare the effectiveness of different diagnostic methods analyzed for the same signal or signals with similar noise complexity.

\subsection{Selectors used for  comparison}

Slices of the spectrogram $\mathbf{S}(f,t)$ are often tested in terms of the occurrence of local fault characteristics. Statistical measures can be used to describe the tightness of the distribution of observed data. Sample kurtosis, as one of the most known data impulsiveness statistics, is chosen to compare IFB selectors. The definition of kurtosis is presented in Section \ref{sec:kurt}. Two other statistics, dedicated to heavy-tailed data, i.e., alpha, from the $\alpha$-stable distribution and conditional variance, are also investigated during IFB selector comparison, where the IFB selector means the characteristic that denotes the informative frequency band. The definitions of the given statistics are presented in Sections \ref{sec:alfa} and \ref{sec:cvb}, respectively.
Another approach uses the cyclic characteristic of a local fault and investigates the existence of dependencies with the spectrogram slices to find the IFB. The most well-known measure of dependence is the Pearson correlation coefficient, described in Section \ref{pea_sel}.

The presented selectors are applied to the entire signal $\mathbf{x}$ of length $N$.

\subsubsection{Spectral kurtosis}
\label{sec:kurt}
The kurtosis statistic is the most known impulsive measure in probability and statistical theory, see \cite{kurt}. It gives the knowledge of the tails of the signal. If the signal has a Gaussian distribution, the statistic is equal to 0. Let consider the spectrogram $\mathbf{S}=[\mathbf{s}_{f_1},...,\mathbf{s}_{f_F}]$ as the set of vectors, then the sample kurtosis for the vector $\mathbf{s}_{f_k}$ is defined as follows \cite{emp_kurt}:

\begin{eqnarray}
\hat{K}(\mathbf{s}_{f_k})=\frac{\frac{1}{T_1}\sum_{i=1}^{T_1}\left(s_{f_ki}-\overline{\mathbf{s}}_{f_k}\right)^4}{\left(\frac{1}{T_1}\sum_{i=1}^{T_1}\left(s_{f_ki}-\overline{\mathbf{s}}_{f_k}\right)^2\right)^2} -3,
\end{eqnarray} 
where $\overline{\mathbf{s}}_{f_k}$ is the sample mean and $T_1$ is the sample length of $\mathbf{s}_{f_k}=[s_{f_k1},\dots,s_{f_k{T_1}}]$ which is calculated for the entire signal $\mathbf{x}$. It is the most frequently used sparsity index in the diagnosis of bearing faults \cite{antoni2006spectral} used for IFB selection. Kurtosis  $\hat{K}(\cdot)$ applied to the slices of the spectrogram $\hat{K}(\mathbf{S})$  is commonly known as the spectral kurtosis \cite{antoni2006spectral}. In the considered paper it is used for comparison of IFB selectors.

It is known that methods, that are based on the kurtosis statistic, have some limitations. The value of the kurtosis decreases when the repetition rate of the impulses increases. If the cyclic impulses become so frequent that they overlap over each other, then the kurtosis vanishes \cite{pachaud1997crest}. On the other hand, when the outliers (non-cyclic impulses) are observed in the recorded signal then the kurtosis takes maximal value. This issue was addressed in \cite{BARSZCZ2011431}.

\subsubsection{Alpha selector}
\label{sec:alfa}
The Gaussian distribution is a fundamental distribution that is used throughout science, however, there are phenomena in which it is not appropriate representation. It happens especially when extreme values are observed in the data. Therefore, a more general distribution is needed that handles the behavior of outlier values.
The possible extension of the Gaussian distribution is $\alpha$ -- stable distribution, see \cite{Taqqu}. The fundamental parameter of this distribution is the stability index $\alpha\in(0,2\rbrack$, which indicates the distance from the Gaussian distribution. For the $\alpha=2$ the $\alpha$ -- distribution simplifies to the Gaussian distribution with some parameters $\mu, \sigma$. If $\alpha$ tends to $0$ the examined distribution becomes more impulsive, that is, the values of outliers and their amplitudes increase. The random variable $X$ has an $\alpha-$ stable distribution if its characteristic function is defined as \cite{Taqqu}:
\[
\mathbb{E}[\exp{i\theta X}]=\phi_X(\theta)=
\begin{cases}
e^{-\sigma^{\alpha}|\theta|^{\alpha}\left\{1-ib \mathrm{sign}(\theta)\tan\left(\pi\alpha/2\right)\right\}+i\mu \theta}, & \alpha\neq 1,\\
e^{-\sigma|\theta|\{1+ib \mathrm{sign}(\theta)\frac{2}{\pi}\log(|\theta|\}+i\mu \theta}, & \alpha= 1.
\end{cases}
\]
The parameter $\sigma>0$ is responsible for the scale, $b \in [-1,1]$ for the skewness, and $\mu \in \mathbb{R}$ is a shift parameter.
To estimate the parameter $\alpha$ from the $\alpha-$stable distribution one can use the McCulloch method \cite{cul}, which uses the following definition:

\begin{eqnarray}
\hat{\alpha}(\mathbf{s}_{f_k})=l(\hat{v}_\alpha,\hat{v}_b), \, \hat{v}_\alpha=\frac{\hat{\mathbf{s}}_{f_k,0.95}-\hat{\mathbf{s}}_{f_k,0.05}}{\hat{\mathbf{s}}_{f_k,0.75}-\hat{\mathbf{s}}_{f_k,0.25}},\, \hat{v}_{b}=\frac{\hat{\mathbf{s}}_{f_k,0.95}-\hat{\mathbf{s}}_{f_k,0.05}-2\hat{\mathbf{s}}_{f_k,0.5}}{\hat{\mathbf{s}}_{f_k,0.95}-\hat{\mathbf{s}}_{f_k,0.05}},
\end{eqnarray}
where $\hat{\mathbf{s}}_{f_k,z}$ is the sample quantile of order $z$ based on the vector $\mathbf{s}_{f_k}=[s_{f_k1},\dots,s_{f_kT_1}]$, $T_1$ is the sample length of $\mathbf{s}_{f_k}$ which is calculated for the entire signal $\mathbf{x}$. The parameter $\hat{\alpha}$ is obtained by linear interpolation $l(\cdot,\cdot)$ of the $v_\alpha$ and $v_b$ bases on the McCulloch tabulation of the function $l(\cdot,\cdot)$, see Table III in \cite{cul}. The scale and location parameters, $\sigma$ and $\mu$, can be estimated in a similar way.
The use of the parameter $\alpha$ can be found in \cite{Yu2013155, HebdaSobkowicz2020}. The IFB selector based on the spectrogram $\mathbf{S}$ and $\alpha$ parameter is defined as the $2-\hat{\alpha}$ and called the Alpha selector. 
The Alpha selector was found to be an effective tool for the selection of IFB in bearing fault diagnostics, especially in the case of non-Gaussian impulsive background noise. As mentioned, if the amplitude of the impulses in the examined data distribution increases, then the $\hat{\alpha}$ parameter tends to $0$ and the Alpha selector increases.

\subsubsection{Conditional variance based selector}
\label{sec:cvb}

Taking into account the growing demand for tools that work properly under non-Gaussian noise conditions, a conditional variance-based selector, dedicated to signals with non-Gaussian noise, has appeared in the literature \cite{Hebda-Sobkowicz2020mssp}.
 The idea of a conditional variance statistic originates from the statistical phenomenon commonly known as the $20/60/20$ rule \cite{pit2016,pit2019}. 
This rule says that if the population is divided into three groups, that is, $20\%$ of the smallest, $60\%$ of the middle, and $20\%$ of the largest values, then the given groups reveal some kind of balance, and their variances are equal. In the paper \cite{pit2016}, the mathematical illustration that justifies this rule in many real-world situations has been demonstrated. It has also been shown that, for any population that can be described by a multidimensional normal vector, this fixed ratio leads to a global equilibrium state.

For the purpose of signals with heavy-tailed distributed noise, partitioning into 7 groups $A_i$ was proposed. The estimators of $A_i$ are defined as follows \cite{pit2019}:
\begin{equation}\label{AA}
    \begin{aligned}
\hat{A}_1 &:=(-\infty,~\hat{\mathbf{s}}_{f_k,0.004}],\\
\hat{A}_2 &:=(\hat{\mathbf{s}}_{f_k,0.004}, ~\hat{\mathbf{s}}_{f_k,0.062}],\\
\hat{A}_3 &:=(\hat{\mathbf{s}}_{f_k,0.062},~\hat{\mathbf{s}}_{f_k,0.308}],\\
\hat{A}_4 &:=(\hat{\mathbf{s}}_{f_k,0.308},~\hat{\mathbf{s}}_{f_k,0.692}],\\
\hat{A}_5 &:=(\hat{\mathbf{s}}_{f_k,0.692},~\hat{\mathbf{s}}_{f_k,0.938}],\\
\hat{A}_6 &:=(\hat{\mathbf{s}}_{f_k,0.938},~\hat{\mathbf{s}}_{f_k,0.996}]\\
\hat{A}_7 &:=(\hat{\mathbf{s}}_{f_k,0.996},~\infty),
    \end{aligned}
\end{equation}
where  
$\hat{\mathbf{s}}_{f_k,z}$ is the empirical quantile of order $z$ calculated for vector $\mathbf{s}_{f_k}$. The conditional variance statistic used in \cite{Hebda-Sobkowicz2020mssp} for the diagnosis of bearing faults is defined as follows:
\begin{eqnarray}\label{stat_N2}
\widehat{CV}(\mathbf{s}_{f_k})
:=\left(\frac{\hat{\sigma}^2_{A_3}-\hat{\sigma}^2_{A_4}}{\hat{\sigma}}+\frac{\hat{\sigma}^2_{A_5}-\hat{\sigma}^2_{A_4}}{\hat{\sigma}}\right)^2\sqrt{T},
\end{eqnarray}
where $\hat{\sigma}_{A_i}$ denotes the estimator of the standard deviation $\sigma_{A_i}$ in the given set $A_i$.
The lower index 7 of $A_{7}$ refers to the number of partitions in which the distribution of the vector $\mathbf{s}_{f_k}=[s_{f_k1},\dots,s_{f_kT}]$ has been divided.  The main property of the divisions $A_i$ is that their variances are equal. Assuming the Gaussian distribution, the following equation is fulfilled \cite{pit2019}:
\begin{equation}\label{pom2}
\sigma_{A_1}^2=\sigma_{A_2}^2=\sigma_{A_3}^2=\sigma_{A_4}^2=\sigma_{A_5}^2=\sigma_{A_6}^2=\sigma_{A_7}^2.
\end{equation}
As noted in \cite{pit2016}, the condition in Eq. \eqref{pom2} creates a dispersion balance for conditional populations, and a different number of partitioning sets could be considered. 
After decomposition of the time-frequency signal, the estimator $\widehat{CV}$ applied to the spectrogram $\mathbf{S}$ is called the conditional variance-based selector (CV-based selector). The CV-based selector is capable of distinguishing different impulses that occur based on the distribution of their amplitudes \cite{Hebda-Sobkowicz2020mssp}. However, it does not take into account the periodicity of the impulses. 

\subsubsection{Pearson-based selector}\label{pea_sel}
In \cite{nowicki2021dependency} the authors proposed aggregation of information from the two-dimensional (2D) $\textbf{CM}$ map to the one-dimensional (1D) vector used as filter characteristic, which is called the Pearson-based selector. Aggregation is the average of non-zero values of the matrix $\mathbf{CM}$ {with respect to the first dimension} $j=1,\dots,F$. The correlation map is integrated as follows:
 \begin{equation}
  PE_{SEL}(j) =\frac{\sum\limits_{j=1}^{F}\textbf{cm}_{f_j}\times\mathbb{1}_{\{\textbf{cm}_{f_j}>TH_1\}}}{\#\mathbb{1}_{\{\textbf{cm}_{f_j}>TH_1\}}}, \qquad  j=1,2, \dots,F, 
\label{selector}
\end{equation}
where $\textbf{CM}=[\textbf{cm}_{f_1},\dots,\textbf{cm}_{f_F}]$ is a correlation map based on the Pearson correlation coefficient, defined in Eq. (\ref{eq:sab}). The definition assumes the conditional sum of the correlation values that are greater than the given threshold $TH_1$, determined as the first quartile of the local minima of the empirical density function of $\textbf{cm}_{(f_j)}$ and $\#\mathbb{1}_{\{\textbf{cm}_{f_j}>TH_1\}}$ is the number of values that satisfy the condition $\textbf{cm}_{f_j}>TH_1$ \cite{nowicki2021dependency}.

The IFB selector based on the Pearson correlation map, that is, $PE_{SEL}$ enables data filtration to extract the fault signal. However, during aggregation, the existing imperfection of $\mathbf{CM}$ is highlighted. Therefore, a threshold $TH_2$ has been proposed to improve its selectivity. The last step is to enhance the selector to remove the values of the selector that correspond to the noise. The threshold $TH_2$ has been defined as the third quartile (Q3) of the selector values multiplied by $1.1$.

 Value 1.1 has been arbitrarily established as a result of the analysis of the simulated signals. Finally, the selector values are as follows:
 \begin{equation}
  PE_{SEL}(j) = \begin{cases}
        PE_{SEL},  & \text{if } PE_{SEL}>TH_2, \\
        0, & \text{otherwise.} 
    \end{cases} 
\label{selector1}
\end{equation}
Thresholding has been proposed to improve its selectivity. However, the proposed threshold is not universal, and a new procedure is needed.

All considered selectors are normalized to facilitate comparison of different selectors.

\section{Simulations} \label{simul}
\subsection{Model of the signal and preliminary analysis}
The synthetic signal considered here mimics a real signal that was encountered in the raw material industry during the measurement of bearing vibrations in machines, such as a crusher or a sieving screen. The signal model $x$ consists of three components, namely a cyclic impulsive signal (with a cycle related to the frequency of failure of the bearings), called an SOI with amplitude $A_{CI}$, random impulses (at random location $t_{NCI}$ and random amplitude $A_{NCI}$), called $IMP_{noise}$ and Gaussian noise with variance $\sigma_{noise}$, that is:

$$x=SOI+IMP_{noise}+N(0,\sigma_{noise}).$$ 

 {Cyclic impulses that constitute the SOI consist of short term harmonic oscillation with damping described by \cite{Wodecki2020}:}
\begin{equation}
    SOI(t)= A_{CI}\cdot\sin{(2 \pi f_c t)}e^{-dt},
\end{equation}
 {where $A_{CI}$ is the amplitude, $t$ is time, $f_c$ is the center frequency of the carrier band of the impulse and $d$ is a decay factor for the exponential function. Note that non-cyclic impulses are designed in the same way, however, $A_{NCI}$ is random, and center frequency $f_c$ is defined separately for the simulated faults and random impulse.}

\begin{figure}[h!]
    \centering  \includegraphics[width=0.8\linewidth]{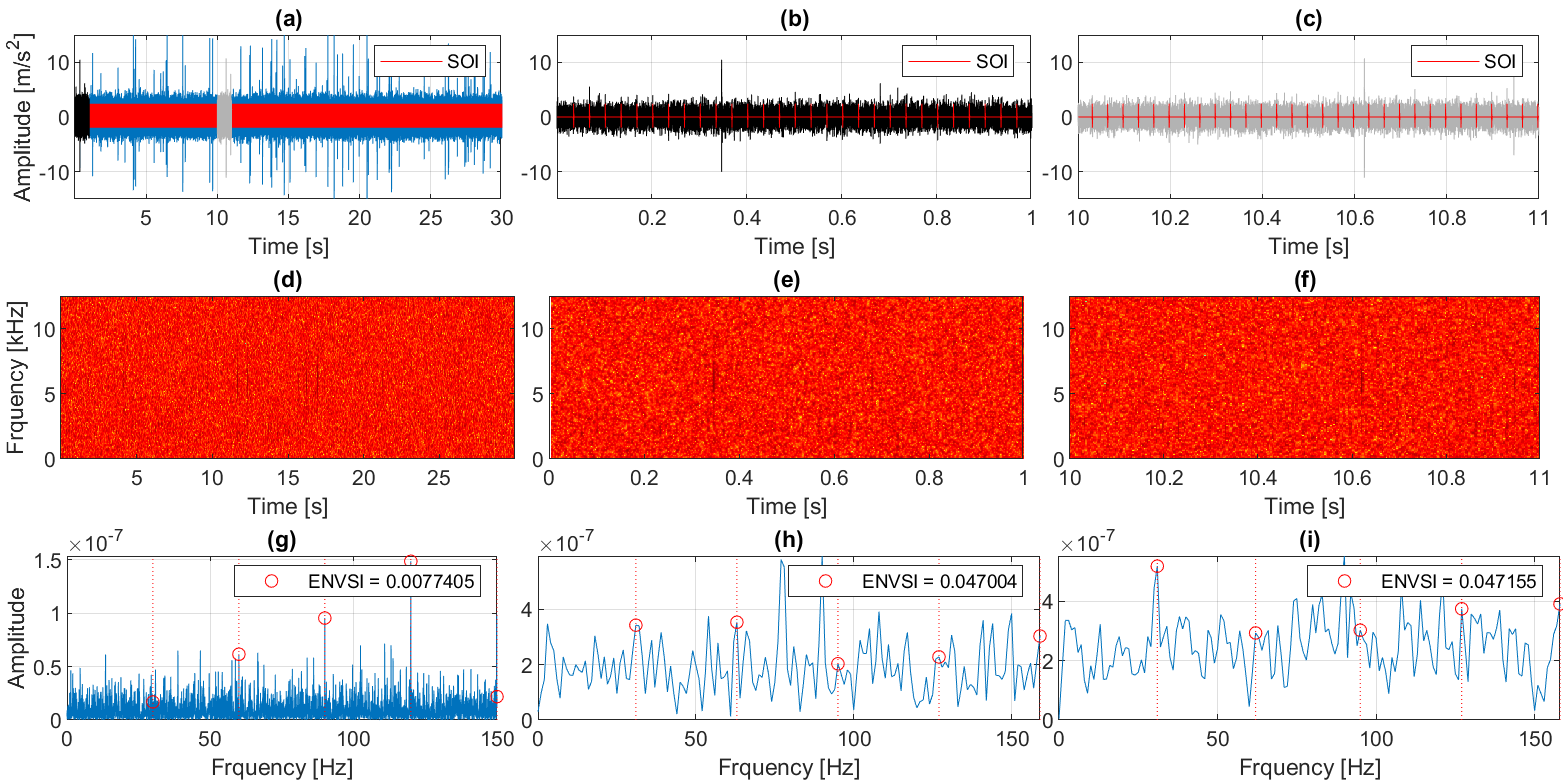}
    \caption{{Simulated signal 30 second length (a) with 1 second length signal fragments presented in (b) and (c), its spectrograms (d) - (f) and SES (g) - (i) respectively.}}
    \label{fig:preliminary_simul}
\end{figure}

 For the purpose of the paper, the following parameters were assumed: $A_{CI}=4$, $A_{NCI}=20$, and $\sigma_{noise}=1.2$.  {Center frequencies for cyclic and non-cyclic impulses are 2500 and 6000 Hz, respectively.}
 The effectiveness of fault detection procedures is significantly dependent on the relationship between $A_{CI}$, $A_{NCI}$, and $\sigma_{noise}$.
Values of these parameters have been selected to imitate real signals with the SOI completely hidden in the noise with randomly spaced, high-amplitude, non-cyclic impulses. The amplitude of non-cyclic impulses is several times higher than the amplitude of background noise. Such conditions are typical for the normal operation of crushers or screens in the raw material industry. Falling pieces of rock excite natural frequencies, and in this way, an impulsive disturbance is created. The number, location, and size of random impulses depend on the granulation of the material. As the location of non-cyclic impulses is random (see Fig. \ref{fig:preliminary_simul} a), it may happen that, for example, one second of the signal will cover a single impulse or even will not contain any non-cyclic impulses (see Fig. \ref{fig:preliminary_simul} b, c). Unfortunately, due to the high variance of Gaussian noise, cyclic impulses are still hidden and cannot be detected neither visually on the spectrogram nor in the ES, see Fig. \ref{fig:preliminary_simul} d-i). 

\subsection{Results for simulated signals} 
 {In this section, the step-by-step partial results for simulated signals are presented. The simulated signal presented in Fig. \ref{fig:preliminary_simul} has been divided into $M=30$ segments. {The division into M segments is a compromise between the length of the segment (to achieve optimal signal resolution) and the largest possible number of segments. The number of segments is crucial in the NTF decomposition to effectively separate the informative and non-informative components. The segment length should contain at least several full periods (usually not less than three) of the tested fault frequency.} For each segment, a spectrogram has been calculated. Then, a dependence map was calculated for each spectrogram. In this case, the dependence map is the Pearson correlation for each pair of frequency bins for each spectrogram.}

\begin{figure}
    \centering
\includegraphics[width=0.69\linewidth]{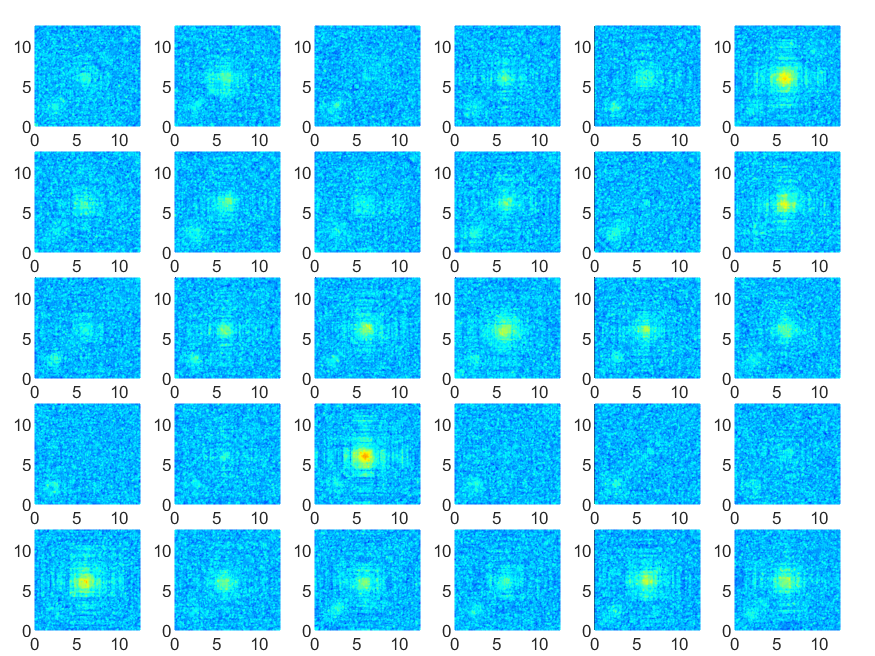}
    \caption{Pearson correlation maps for the simulated signal.}
    \label{fig:pearson_sym_maps}
\end{figure}

\begin{figure}[h!]
     \centering
     \begin{subfigure}[b]{0.4\textwidth}
         \centering
         \includegraphics[width=\textwidth]{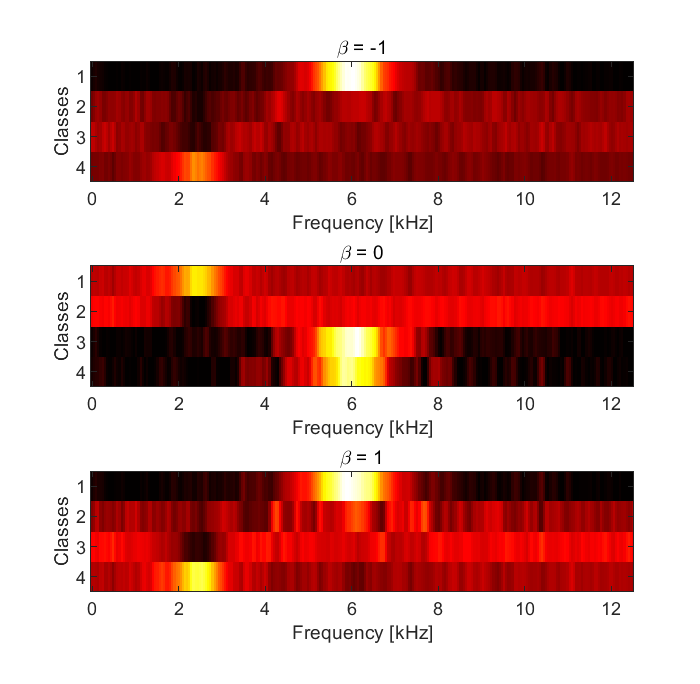}
         \caption{Matrix H  from NTF decomposition}
         \label{fig:xx}
     \end{subfigure}
     \begin{subfigure}[b]{0.4\textwidth}
         \centering
         \includegraphics[width=\textwidth]{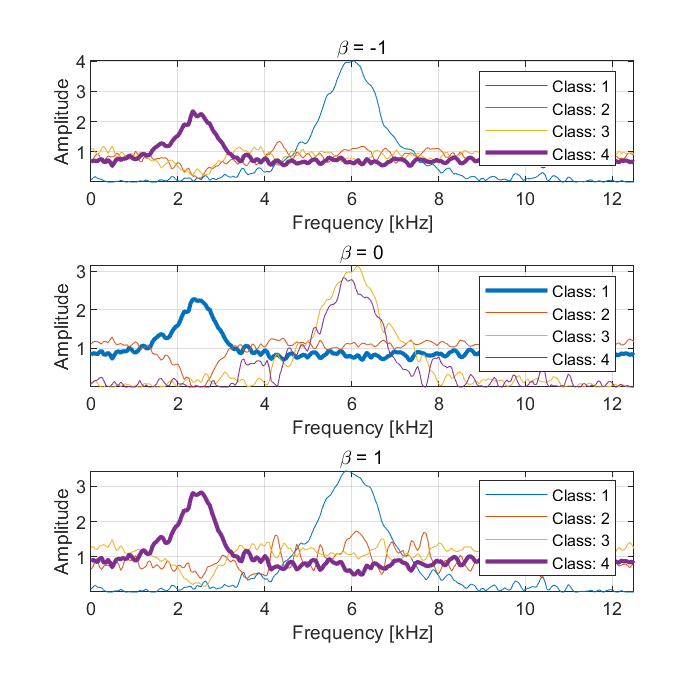}
          \caption{Selectors extracted from matrix H}
         \label{fig:xx}
     \end{subfigure}
    
        \caption{{NTF decomposition results: matrix $\bold{H}$, $N=30$, simulated signal: (a) feature vectors in $\bold{H}$ for all classes (b) Selectors based on feature vectors in $\bold{H}$ for all classes. The selector corresponding to the IFB is shown in a bold line.}}
        \label{fig:results_simul}
\end{figure}

 All dependence maps (Pearson correlation matrices) for 1 second length signals are presented in Fig. \ref{fig:pearson_sym_maps}. All maps are presented on the same color scale.  One may easily note that some maps are different in the sense of the structure of dependence. In the next step they are taken for the NTF analysis. As one can notice, the visibility of the desired signal in the band of 2-3kHz (related to the cyclic impulses) depends on the signal fragment.
The results obtained for the proposed method are presented in Fig. \ref{fig:results_simul}. The left panel displays the matrices $\bold{H}$ for three values of $\beta$. The column vectors in $\bold{H}$ contain the frequency features that are illustrated as the rows (four here) in the presented graphs. {In the analyzed case, it was assumed that the number of classes corresponds to the components expected to be encountered while analyzing such a signal. The four classes include damage, non-cyclic pulses, noise, and an additional one (which does not align with the other three identified classes).} The frequency features can be interpreted as selectors of some specific information. The number of vectors is related to the complexity of the spectral structure (a number of sources in the observed signal). The right panel is more intuitive from a local damage perspective, where the same information is presented in the form of an amplitude-frequency characteristic that is directly the shape of the filter.
In Fig. \ref{fig:results_simul} b, one can recognize two bands with higher amplitudes indicated by different curves. The 2-3 kHz band is informative (related to cyclic impulses), and the band around 6 kHz results from a natural resonance excited by pieces of material hitting the housing of a crushing machine (non-cyclic impulses). It is worth highlighting that the characteristics indicating the informative band are very selective and could be easily thresholded and normalized to achieve an almost perfect filter.
\begin{figure}[h!]
    \centering    \includegraphics[width=0.7\linewidth]{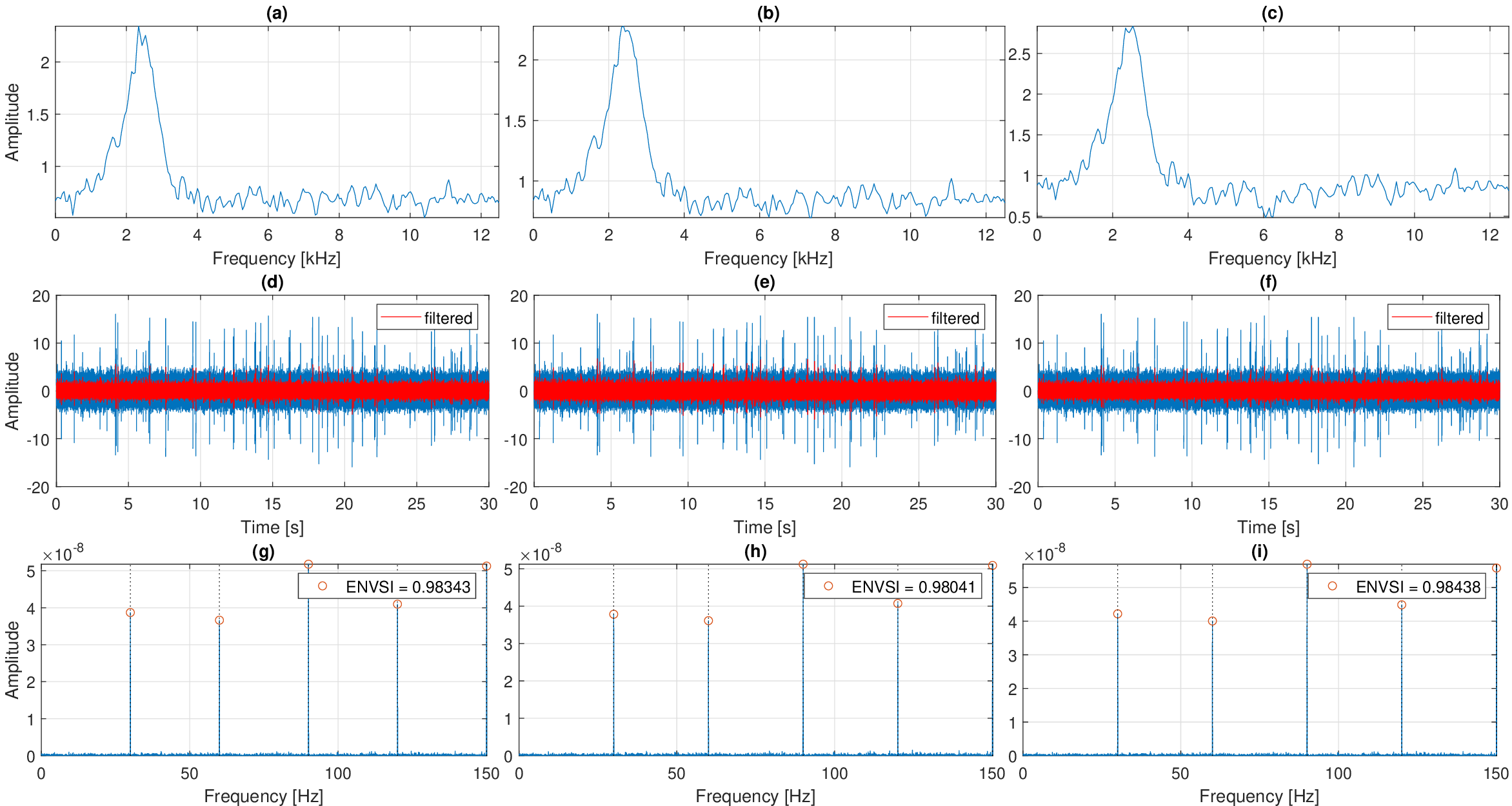}
   \caption{Filter characteristics: (a) NTF $\beta=-1$ (b) NTF $\beta=0$ (c) NTF $\beta=1$ for the simulated signal, filtration results (d) - (f), and SES of the filtered signals with given selectors (g) - (i), respectively}
\label{fig:final_results_simul}
\end{figure}
{The filter characteristics presented in Fig. \ref{fig:final_results_simul}, which indicate the band of 2-3 kHz, are used to pre-filter the signal.} The raw (blue) and filtered (red) signals are presented in second row, subplots (d) - (f). The squared envelope spectrum and the calculated ENVSI value are presented in the third row of pictures (subplots (g) - (i)). ENVSI values are very high (close to 1), which means that the damage is detectable in a relatively easy way.

\subsection{Comparison with popular selectors}
\label{com_sim}
\begin{figure}[h!]
    \centering
    \includegraphics[width=0.8\linewidth]{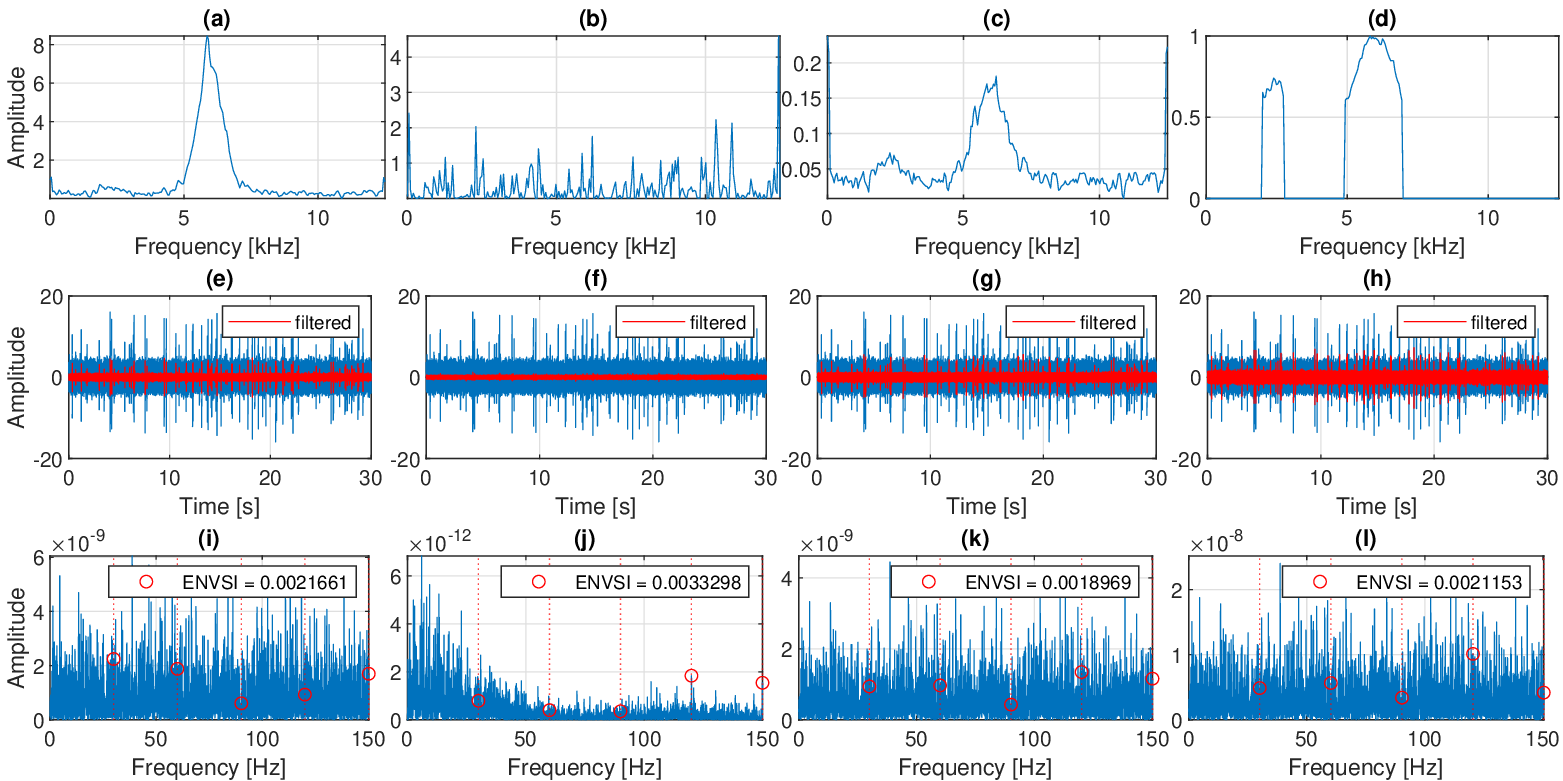}
    \caption{Selectors results: (a) spectral kurtosis, (b) CV, (c) alpha, (d) Pearson for the simulated signal $AC=4$, $ANCI=20$, filtration results (e) - (h), and SES of the filtered signals with given selectors (i) - (l), respectively }
    \label{fig:comparison_simul}
\end{figure}

In order to validate the proposed approach, received results have been compared with the results obtained from recently developed popular selectors. In Fig. \ref{fig:comparison_simul} the results for four selectors are presented, namely spectral kurtosis, CV-based selector, Alpha selector, and Pearson-based selector (see Figs. \ref{fig:comparison_simul} a-d). None of the first three selectors indicates the expected informative frequency band (2-3kHz). Even if the Pearson-based selector identifies this band (see Fig. \ref{fig:comparison_simul} d), it also indicates the band related to non-cyclic impulses (6 kHz). As non-cyclic impulses have significantly higher amplitudes, they will also dominate the signal after filtering.
The second row of the pictures in Fig. \ref{fig:comparison_simul} shows the original observation (blue) and the signals extracted using estimated filters (see Figs. \ref{fig:comparison_simul} e-h). The extracted signals are very weak and do not reveal a clear cyclic impulsive behavior. Figs. \ref{fig:comparison_simul} a, c, and d show that the signal after filtering should contain non-cyclic impulses in the band around 6kHz. The filter illustrated in Fig. \ref{fig:comparison_simul} b does not select anything. Finally, the squared envelope spectrum is plotted for the extracted signals to search (detect) a family of fault frequencies (see Figs. \ref{fig:comparison_simul} i - l).
As can be seen, any component of the fault frequency cannot be practically detected, and the ENVSI values are very small, so one may conclude that the selection of the right band using known selectors is not appropriate.

\begin{figure} [h!]
     \centering
         \centering
         \includegraphics[width=0.8\textwidth]{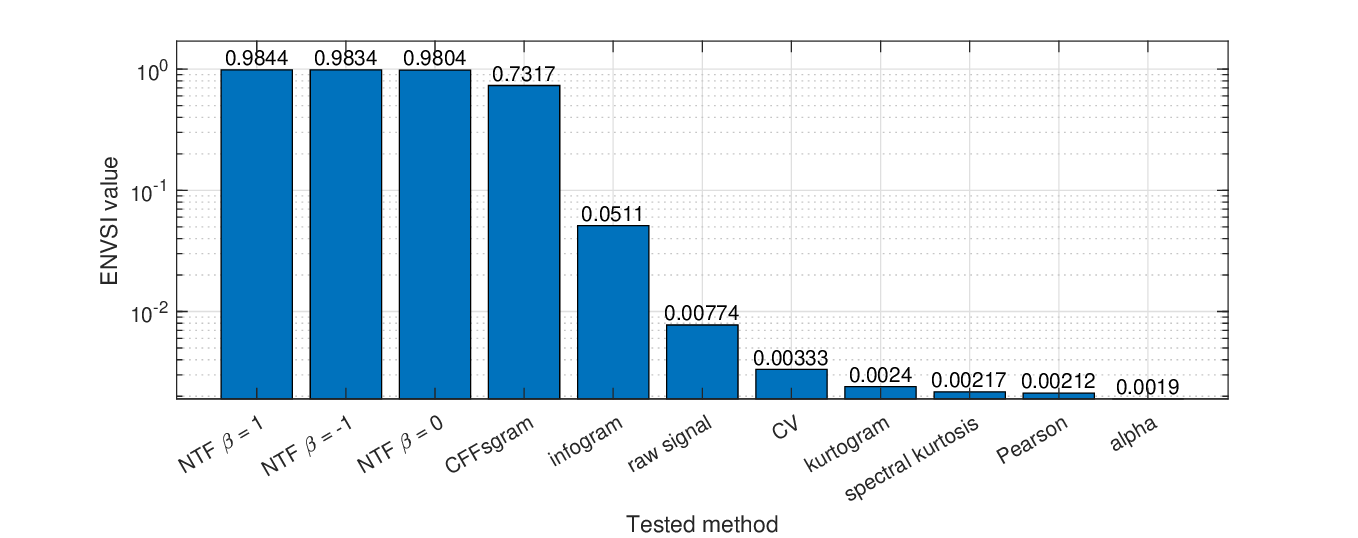}
        \caption{{Comparison of sorted ENVSI values of the simulated signal, filtered signals with the given selectors and filtered according to the kurtogram, Infogram, and CFFsgram results.}}
        \label{fig:envsi_bar_simul}
\end{figure}

{The proposed approach was also compared with the kurtogram \cite{antoni2007fast}, Infogram \cite{antoni2016info}, and CFFsgram \cite{zhou2023cffsgram}. For the simulated data, the results are presneted in Fig. \ref{fig:appendix_simul_gram} in Appendix while SES of the filtered signals (together with the ENVSI value) are demonstrated in Fig. \ref{fig:appendix_simul_SES} in Appendix.  The aggregate results of the ENVSI values, sorted in descending order, are shown in Fig. \ref{fig:envsi_bar_simul}. For better visibility, the indicator values are shown on a logarithmic scale. As can be seen, for all NTF-based selectors, the ENVSI values obtained are higher than 0.98. The CFFsgram allows the fault detection and has the ENVSI equal to 0.7317. Other techniques give values lower than 0.05, so the fault frequency and its harmonics are not visible above the background noise, and fault detection is impossible.}

\section{Real data analysis}\label{real_analysis}
 
 {The approach presented in this manuscript was also tested on two datasets measured in real-life conditions. Case 1 corresponds to the vibration signal from the hammer crusher, while Case 2 corresponds to the acoustic signal measured on the test rig.  }

In most practical situations, it is very difficult to acquire examples of signals from a faulty machine during normal operation in industrial conditions. The considered crusher is a special case because, due to the impulsive character of the load, even early-stage damage can spread quickly and cause catastrophic failure. Therefore, a damaged bearing is a highly undesirable situation. To demonstrate the efficiency of the method, a signal from a healthy bearing installed in the crusher was used and a simulated SOI was added to mimic the faulty component.  It is much easier to model the SOI (as its shape is perfectly known) than to model the background noise of the machine considered.

The vibration signal data used in this section originate from a copper ore crusher operating in the mining industry. The length of the signal is 30 seconds and the sampling frequency is 25 kHz. A local fault has been added with the frequency equal to 30 Hz and the carrier frequency equal to 2.5 kHz ($2-3$ kHz). \textcolor{black}{Similar approach related to the same machine but from another measurement is presented in \cite{wylomanska2016impulsive,HebdaSobkowicz2020,Hebda-Sobkowicz2020mssp,wodecki2021local, hebda2022infogram}}.

\begin{figure}[ht!]
\centering
 \includegraphics[width=0.5\textwidth]{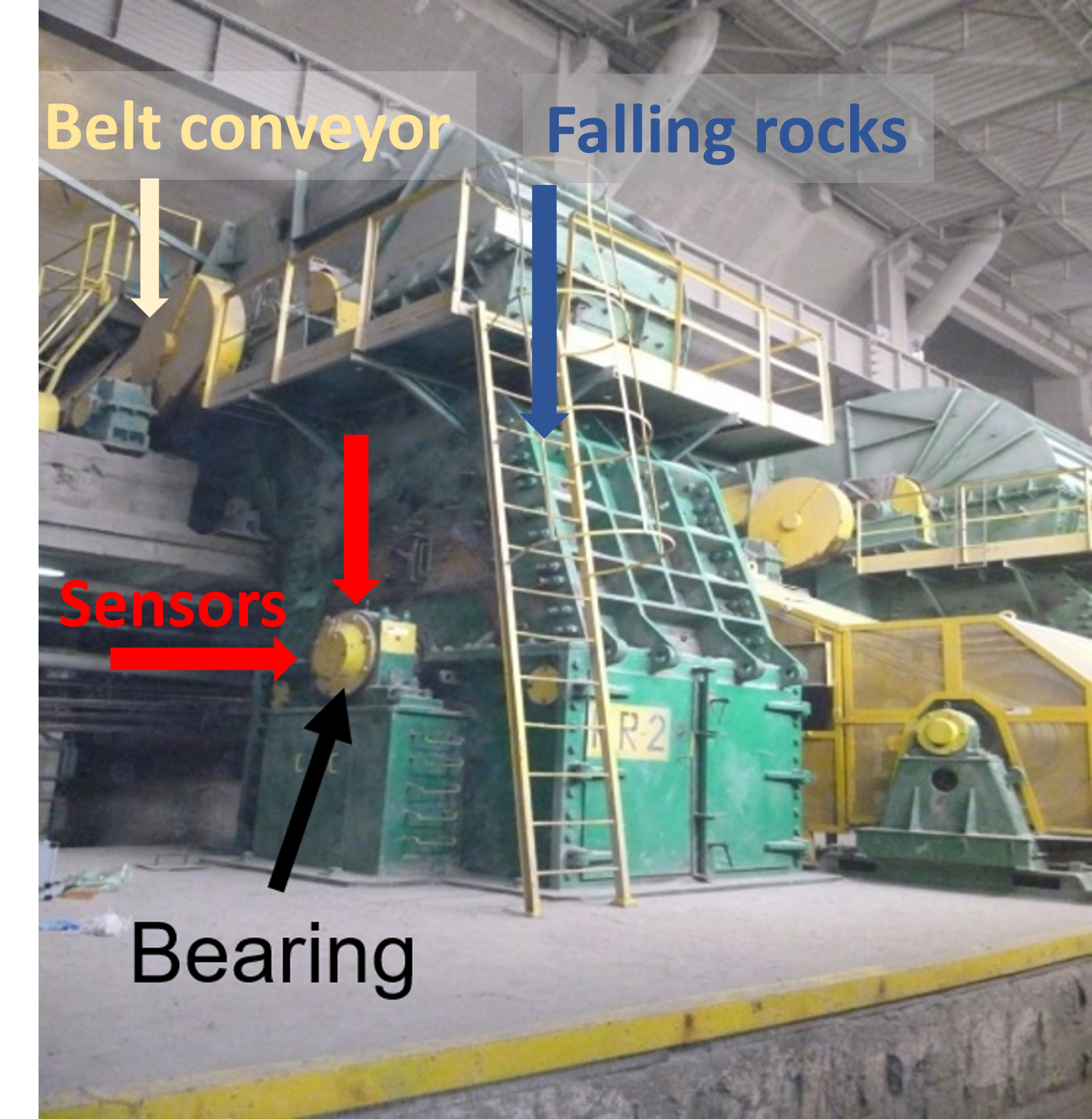}
\caption{Copper ore crusher \textcolor{black}{\cite{wylomanska2016impulsive}}}
\label{fig:obj_crush}
\end{figure}

The signal analyzed in this study is a vibration time series acquired from an ore crusher during a normal operation in the mining industry (see Fig.~\ref{fig:obj_crush}). {A hammer crusher is a machine specifically designed to break down copper ore. Considering the input material stream is crucial because of the presence of copper ore fragments of varying sizes, so the machine is operating under time-varying conditions. The non-Gaussian, impulsive disturbances caused by the larger pieces are observed in the vibration signal. They were modeled in the simulation study as impulsive noise.} The signal has been recorded with a sampling frequency of 25 kHz using Endevco accelerometers and the basic data acquisition system, namely the NI DAQ card and Labview Signal Express software. {Sensors are mounted on the casing of the bearing carrying the main shaft of the machine. This element is indicated in Fig.~\ref{fig:obj_crush} by a black arrow, and the sensors are represented by red arrows. During the analysis, the signal measured with an accelerometer placed in the vertical direction was considered.} The rotational speed during these parts of the signals might be assumed to be approximately constant. The characteristic frequencies of the bearings are presented in Tab. \ref{tab:freqs}. {The bearings considered are 23264 SKF, and one of them reveals local inner race damage. As bolded in Tab. \ref{tab:freqs}, the characteristic frequency of the inner race is equal to 30.7 Hz. } 
\begin{table}[ht!]
  \centering
      \caption{Characteristic frequencies of 23264 CCK/W33 bearing}
  	\resizebox{0.8\textwidth}{!}{
  \begin{tabular}{|l|l|}
  \hline
     \textbf{Description} & \textbf{Value} \\ \hline
     Rotational frequency of the inner ring & 3 Hz\\ \hline
     Rotational freq. of the rolling element and cage assembly & 1.3 Hz\\ \hline
     Rotational freq. of a rolling element about its own axis & 10.6 Hz\\ \hline
     \textbf{Over-rolling frequency of one point on the inner ring} & \textbf{30.7 Hz}\\ \hline
     Over-rolling frequency of one point on the outer ring & 23.3 Hz\\ \hline
     Over-rolling frequency of one point on a rolling element & 21.1 Hz\\ \hline
  \end{tabular}
  }
  \label{tab:freqs}
\end{table}

\begin{figure}[ht!]
\centering
 \includegraphics[width=0.5\textwidth]{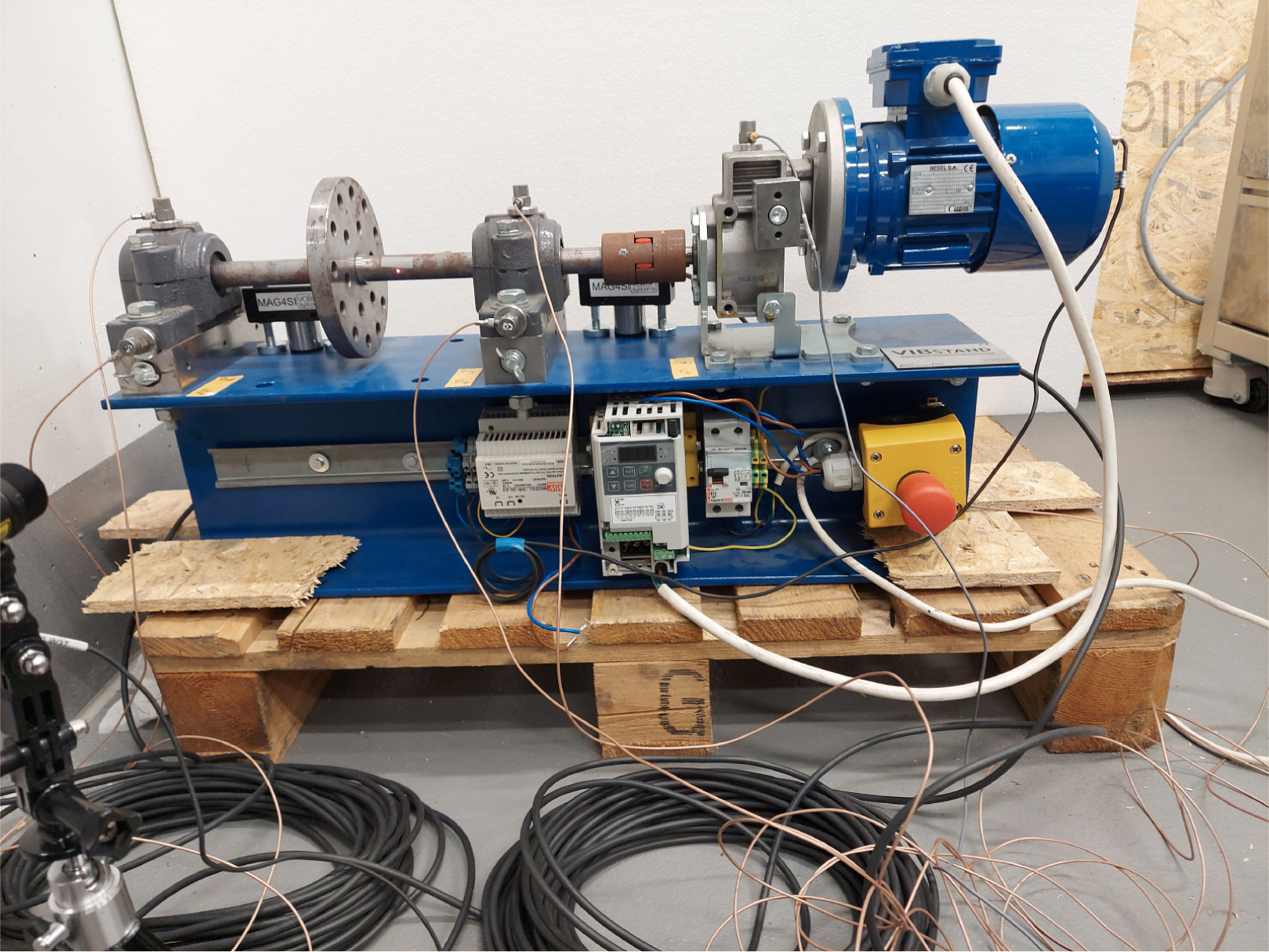}
\caption{Test rig used in the experiment.}
\label{fig:obj_test_rig}
\end{figure}

In Case 2 the signal was obtained using a test rig. The setup included an electric motor, a gearbox, couplings, and two bearings, as presented in Fig \ref{fig:obj_test_rig}. One of the bearings was intentionally damaged to simulate real-world fault conditions. The signal has been recorded with the sampling frequency of 50 kHz using the Bruel \& Kjaer 4189 microphone and the Kistler LabAmp 5165A data acquisition system. The rotational speed was constant during the experiment and equal to 1041 rpm. The characteristic frequencies of the bearings are presented in Tab. \ref{tab:freqs_test_rig}. The bearings considered are 1205 EKTN9 SKF, and one of them reveals local outer race damage. As bolded in Tab. \ref{tab:freqs_test_rig}, the characteristic frequency of the outer race is equal to 91.11 Hz.

\begin{table}[ht!]
  \centering
      \caption{Characteristic frequencies of 1205 EKTN9 bearing}
  	\resizebox{0.8\textwidth}{!}{
  \begin{tabular}{|l|l|}
  \hline
     \textbf{Description} & \textbf{Value} \\ \hline
     Rotational frequency of the inner ring & 17.35 Hz\\ \hline
     Rotational freq. of the rolling element and cage assembly & 7 Hz\\ \hline
     Rotational freq. of a rolling element about its own axis & 42.75 Hz\\ \hline
     Over-rolling frequency of one point on the inner ring & 134.44 Hz\\ \hline
     \textbf{Over-rolling frequency of one point on the outer ring} & \textbf{91.11 Hz}\\ \hline
     Over-rolling frequency of one point on a rolling element & 85.5 Hz\\ \hline
  \end{tabular}
  }
  \label{tab:freqs_test_rig}
\end{table}

\subsection{Preliminary results for real signals}
 
\begin{figure}[h!]
    \centering
    \includegraphics[width=0.8\linewidth]{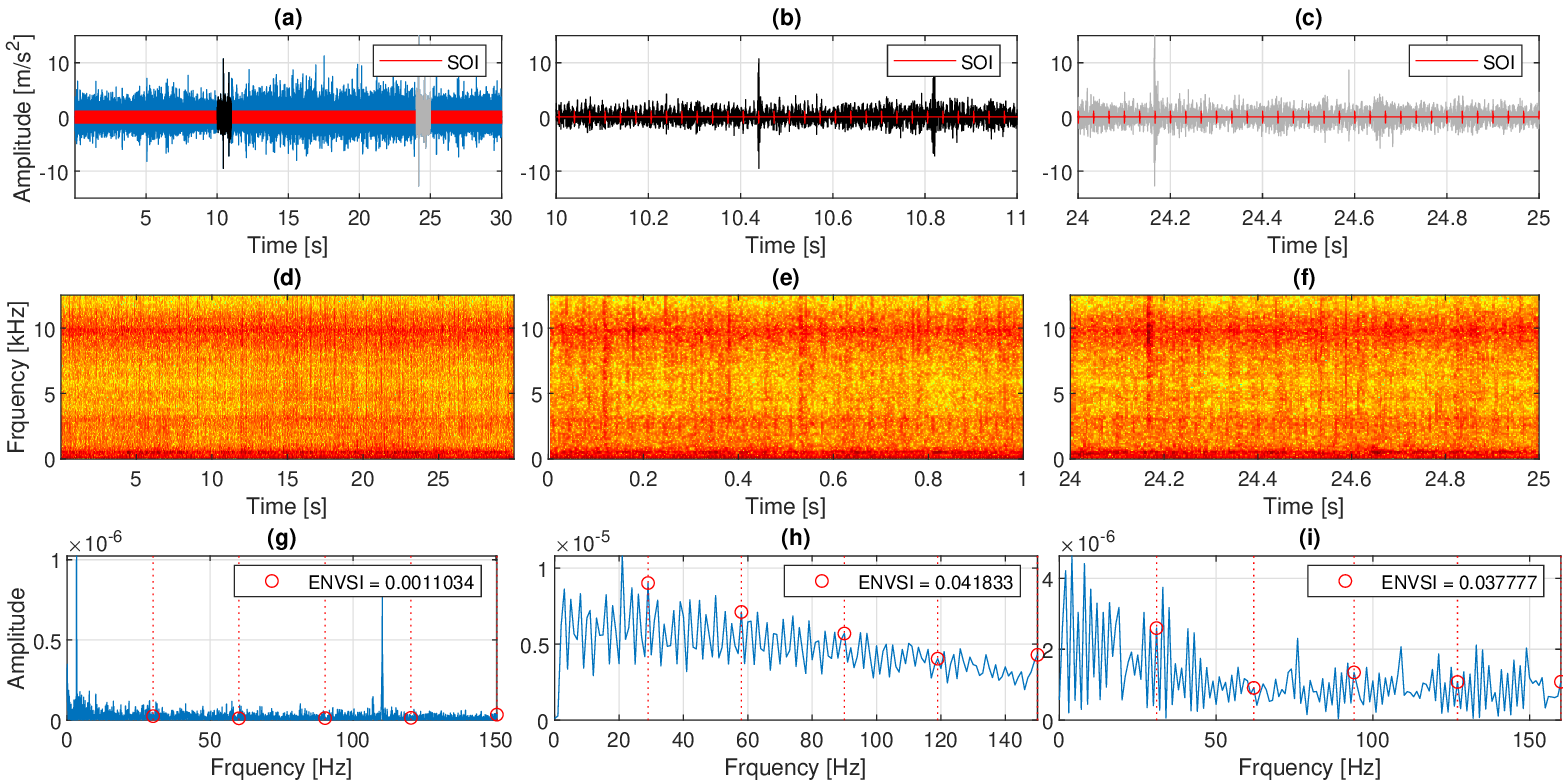}
    \caption{Raw vibration signal  of 30 second length corresponding to the Case 1 (a) with 1 second length signal fragments presented in (b) and (c), its spectrograms (d) - (f), and SES (g) - (i), respectively}
    \label{fig:preliminary_real}
\end{figure}

\begin{figure}[h!]
    \centering
    \includegraphics[width=0.8\linewidth]{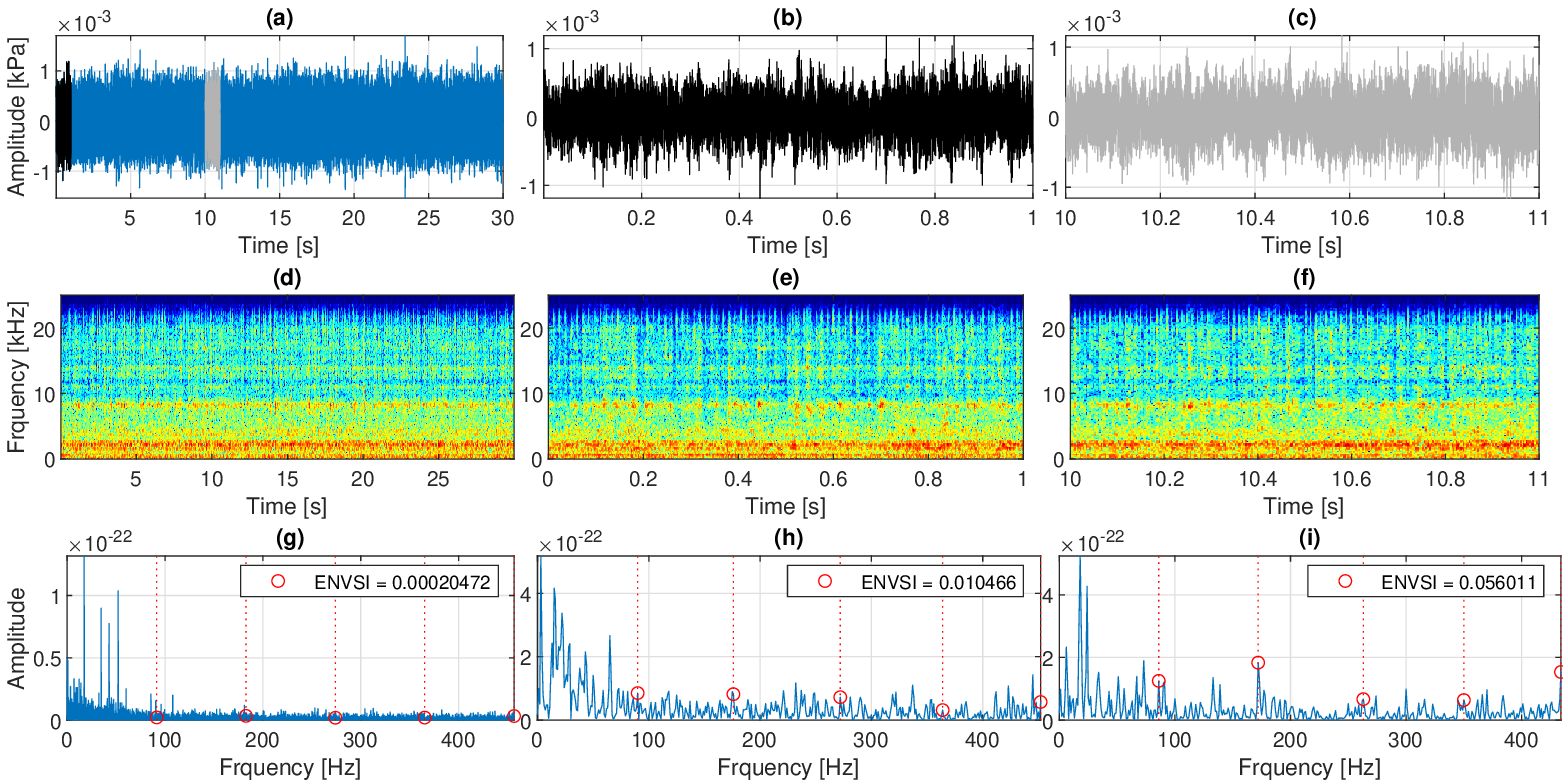}
    \caption{Raw vibration signal of 30 second length corresponding to the Case 2 (a) with 1 second length signal fragments presented in (b) and (c), its spectrograms (d) - (f), and SES (g) - (i), respectively}
    \label{fig:preliminary_real_1}
\end{figure}

The real signals were initially analyzed according to the same convention as the simulation signals. {The raw signal and selected segments for Case 1 are shown in Fig. \ref{fig:preliminary_real} (a-c), while those for Case 2 are presented in Fig. \ref{fig:preliminary_real_1} (a-c). In Case 1, the signal of interest is completely masked by noise; for some segments, the non-cyclic components are several times higher than the averaged amplitude of the signal.
Spectrograms of the entire signal and selected segments are difficult to inspect. The resonance around 10 kHz is clearly seen. In Case 2, the averaged amplitude of the signal is relatively constant. The cyclic component corresponding to the local fault is not visible in the signal or the corresponding spectrograms. Most of the signal energy is concentrated in the low frequencies (0-5 kHz) on the spectrograms. In contrast, the middle and high frequencies display lower energy levels.}

 {{Last (bottom) row of Figs. \ref{fig:preliminary_real_1} and \ref{fig:preliminary_real} presents squared envelope spectra for selected segments of the signal for Case 1 and 2, respectively. In Fig \ref{fig:preliminary_real_1}, one can observe the various components present on the ES; nevertheless, none of them correspond to the fault frequency, which is equal to 91.11 Hz. In Fig. \ref{fig:preliminary_real}}, one should expect a harmonic pattern in the squared envelope spectrum with a fundamental fault frequency equal to 30 Hz. Unfortunately, because of non-Gaussian background noise existing in the signal, informative spectral components are hardly detectable. The ENVSI (envelope spectral indicator), a scalar value that describes the presence of a faulty component in a given signal, has a very small value {in both analyzed cases}.}

\subsection{Results for real signals}

\subsubsection{{Case 1: copper ore crusher}}
\begin{figure}[H]
    \centering
\includegraphics[width=0.69\linewidth]{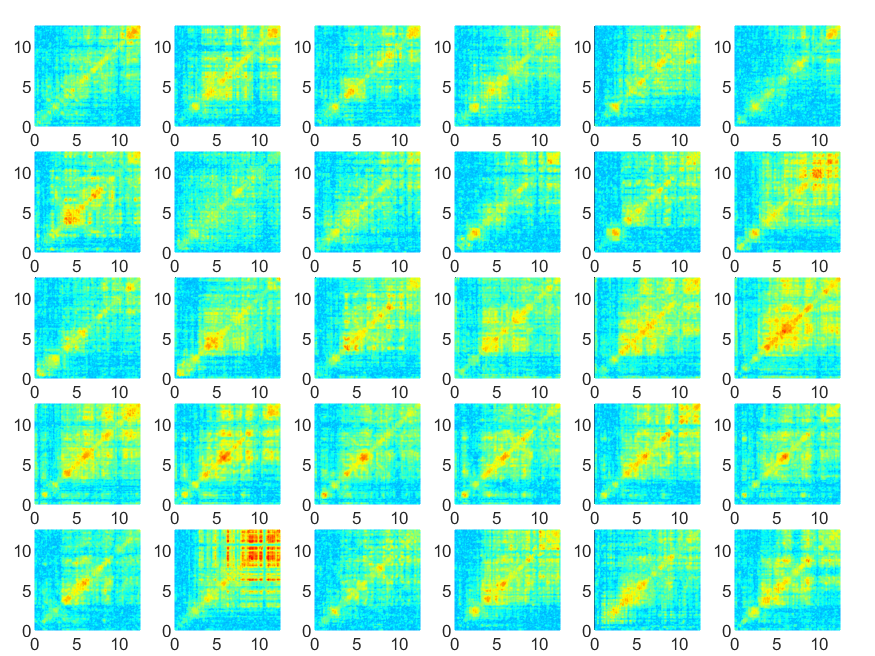}
    \caption{Pearson correlation maps for the real vibration signal {corresponding to Case 1.}}
    \label{fig:pearson_real_maps}
\end{figure}

\begin{figure} [h!]
     \centering
     \begin{subfigure}[b]{0.4\textwidth}
         \centering
         \includegraphics[width=\textwidth]{ 	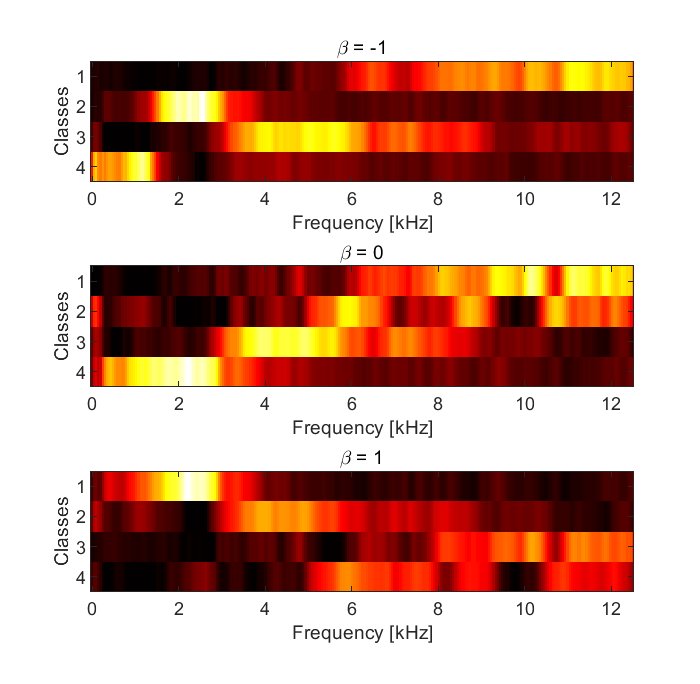}
         \caption{Matrix $\bold{H}$ from NTF decomposition for Case 1}
         \label{fig:xxa}
     \end{subfigure}
     \begin{subfigure}[b]{0.4\textwidth}
         \centering
         \includegraphics[width=\textwidth]{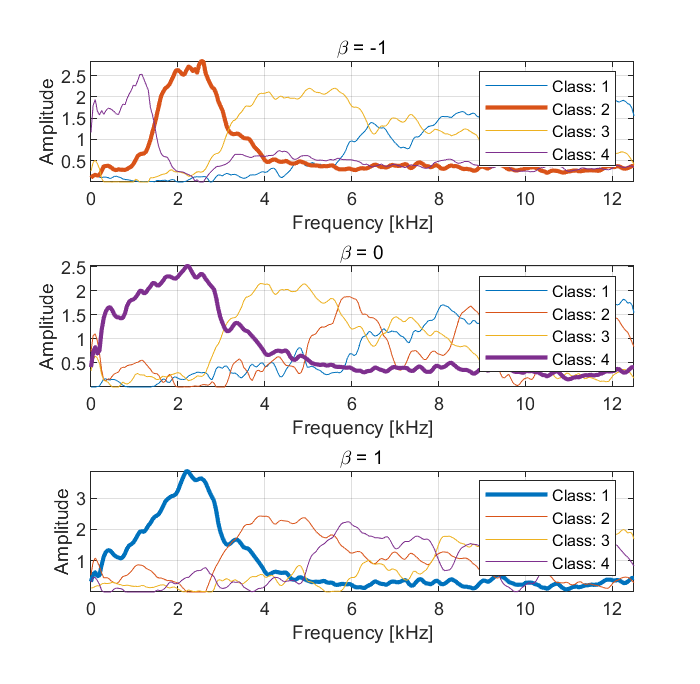}
         \caption{Selectors extracted from matrix $\bold{H}$ for Case 1}
         \label{fig:xxb}
     \end{subfigure}
        \caption{{NTF decomposition results for Case 1: matrix $\bold{H}$, $N=30$, real vibration signal: a) feature vectors in $\bold{H}$ for all classes, b) selectors based on the feature vectors in $\bold{H}$ for all classes. The selector corresponding to the IFB is shown as a bold line.}}
        \label{fig:NTF_real}
\end{figure}

\begin{figure}[h!]
    \centering
    \includegraphics[width=0.8\linewidth]{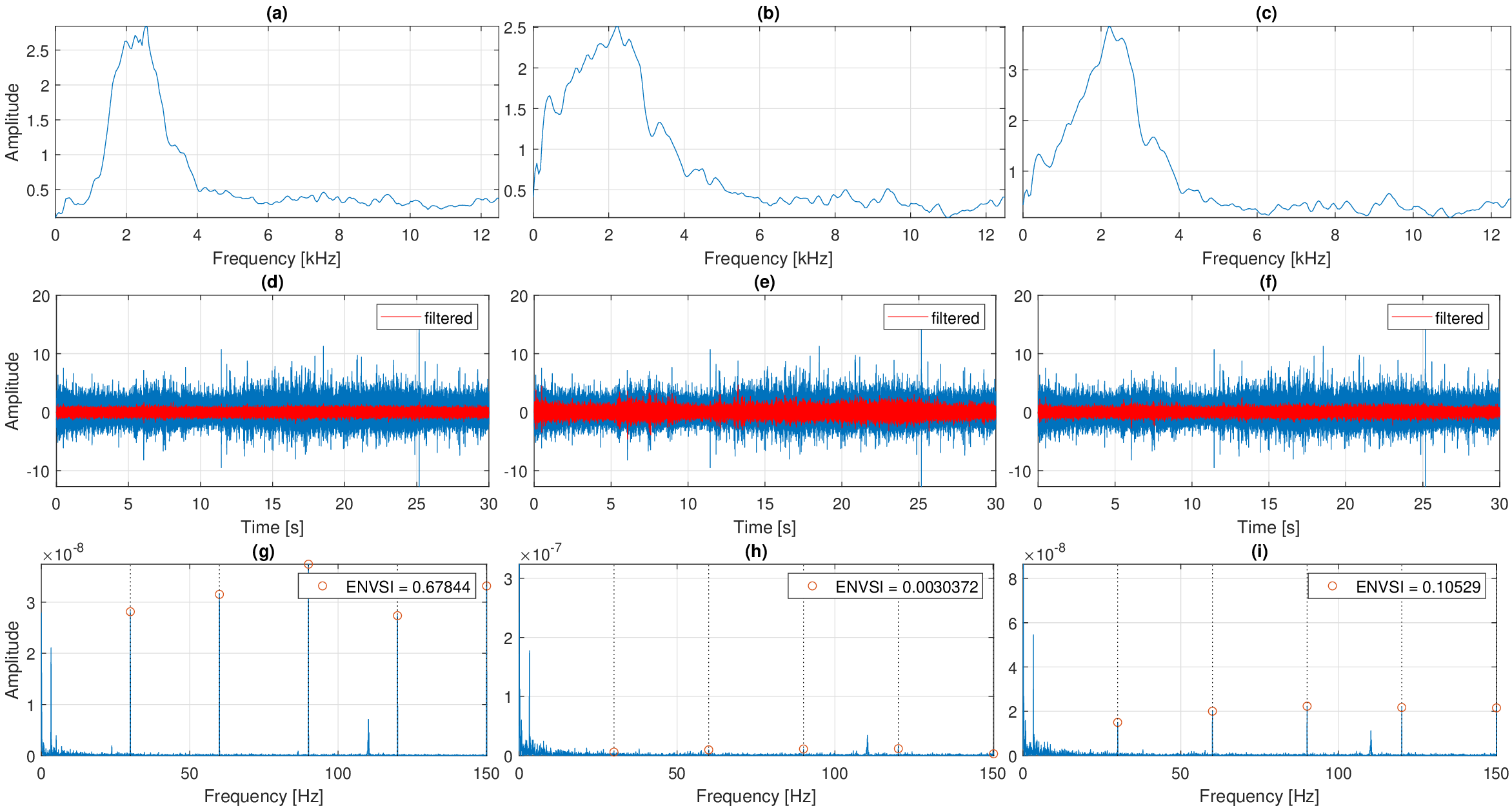}
    \caption{{Selectors results for Case 1: (a) NTF $\beta=-1$ (b) NTF $\beta=0$ (c) NTF $\beta=1$ for the real vibration signal, filtration results (d) - (f), and SES of the filtered signals with  given selectors (g) - (i), respectively.}}
    \label{fig:final_NTF_real}
\end{figure}

The Pearson correlation maps for 1 second length signals are presented in Fig. \ref{fig:pearson_real_maps}. All maps are presented in the same color scale and correspond to {one-second length segments of} 30-second signal used for the NTF analysis. The band corresponding to non-cyclic impulses is dominant in the signal and makes the IFB (2-3 kHz) difficult to see in some maps. This underscores why it is so important to use information from all of them simultaneously.
The results obtained with the proposed method are presented in Fig. \ref{fig:NTF_real}. The left panel shows matrix $\bold{H}$ (with three various values of $\beta$) while the right panel presents the same results in easy to interpret form of the amplitude-frequency characteristics. For each subplot on the right side, one of the characteristics is bold to highlight a filter shape capable of extracting information (the frequency band of 2-3 kHz). The highlighted selector for each value of $\beta$ is very selective indicating an informative band and almost nothing outside of this band.
The selectors obtained with factorization of dependence maps, which are highlighted in Fig. \ref{fig:NTF_real} (right panel, bold line), have been used to filter the original vibration signal from the machine. In Fig. \ref{fig:final_NTF_real} the final results for the given selector characteristics are presented in the same convention for 3 different values of $\beta$.
It is clear that the selector for $\beta=-1$ (see subplot (a)) is the most selective and finally provides the clearest structure of the squared envelope spectrum and the highest ENVSI value (0.678). 
The selector for $\beta=1$ (see subplot (c)) has a bit worse selectivity, however, after transformations to the squared envelope spectrum one may notice significantly smaller amplitudes of components and several times lower value of ENVSI (0.105). 
The selector obtained for $\beta=0$ (see subplot (b)) covers a bit wider band (it is less selective), and the structure of the squared envelope spectrum does not contain informative components (the ENVSI value is very small -- 0.003). It is surprising how small a change in filter selectivity affects the final result.

\subsubsection{{Case 2: test rig}}

\begin{figure} [h!]
     \centering
     \begin{subfigure}[b]{0.4\textwidth}
         \centering
         \includegraphics[width=\textwidth]{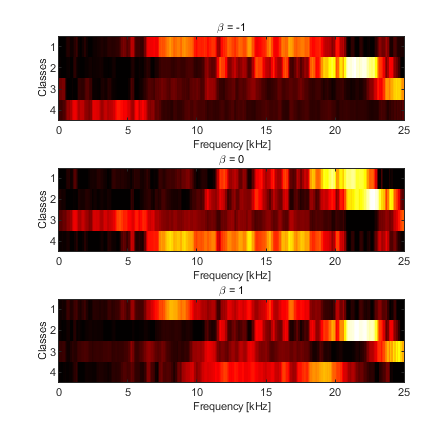}
         \caption{Matrix $\bold{H}$ from NTF decomposition for Case 2}
         \label{fig:xxa}
     \end{subfigure}
     \begin{subfigure}[b]{0.38\textwidth}
         \centering
         \includegraphics[width=\textwidth]{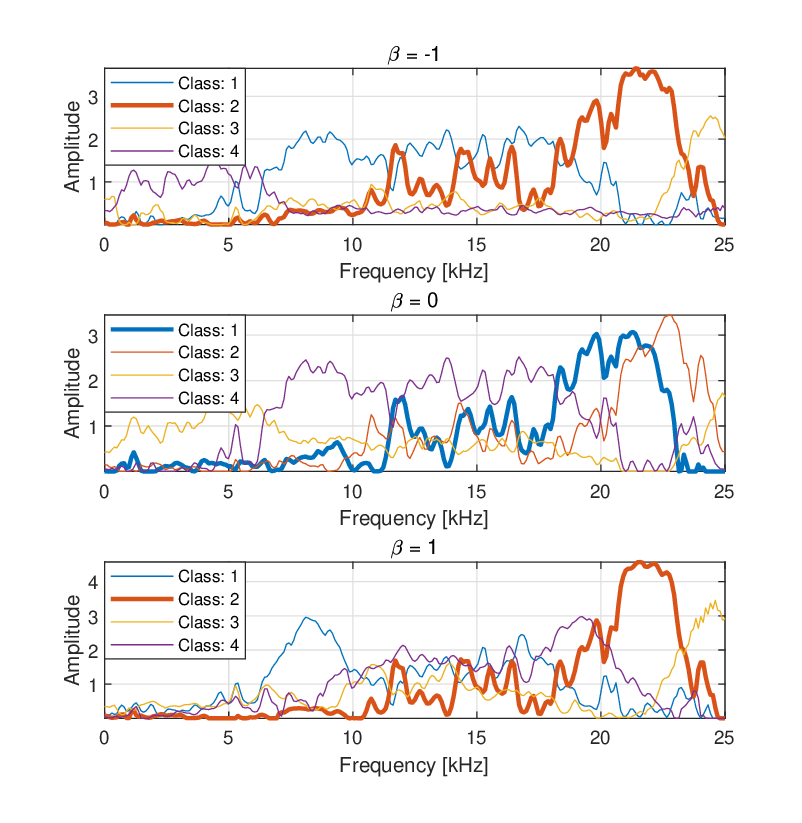}
         \caption{Selectors extracted from matrix $\bold{H}$ for Case 2}
         \label{fig:xxb}
     \end{subfigure}
        \caption{{NTF decomposition results for Case 2: matrix $\bold{H}$, $N=30$, real acoustic signal: a) feature vectors in $\bold{H}$ for all classes, b) selectors based on the feature vectors in $\bold{H}$ for all classes. The selector corresponding to the IFB is shown as a bold line.}}
        \label{fig:NTF_real_2}
\end{figure}

{The results obtained with the proposed approach are presented in Fig. \ref{fig:NTF_real_2}. Similarly, like in Case 1, the left panel displays matrix $\bold{H}$ (with three various values of parameter $\beta$) while the right panel presents the same results in the form of the amplitude-frequency characteristics. In each subplot on the right side, one characteristic is bolded to highlight a filter shape that effectively extracts useful information in the frequency band of 20-24 kHz.}

\begin{figure}[h!]
    \centering
    \includegraphics[width=0.8\linewidth]{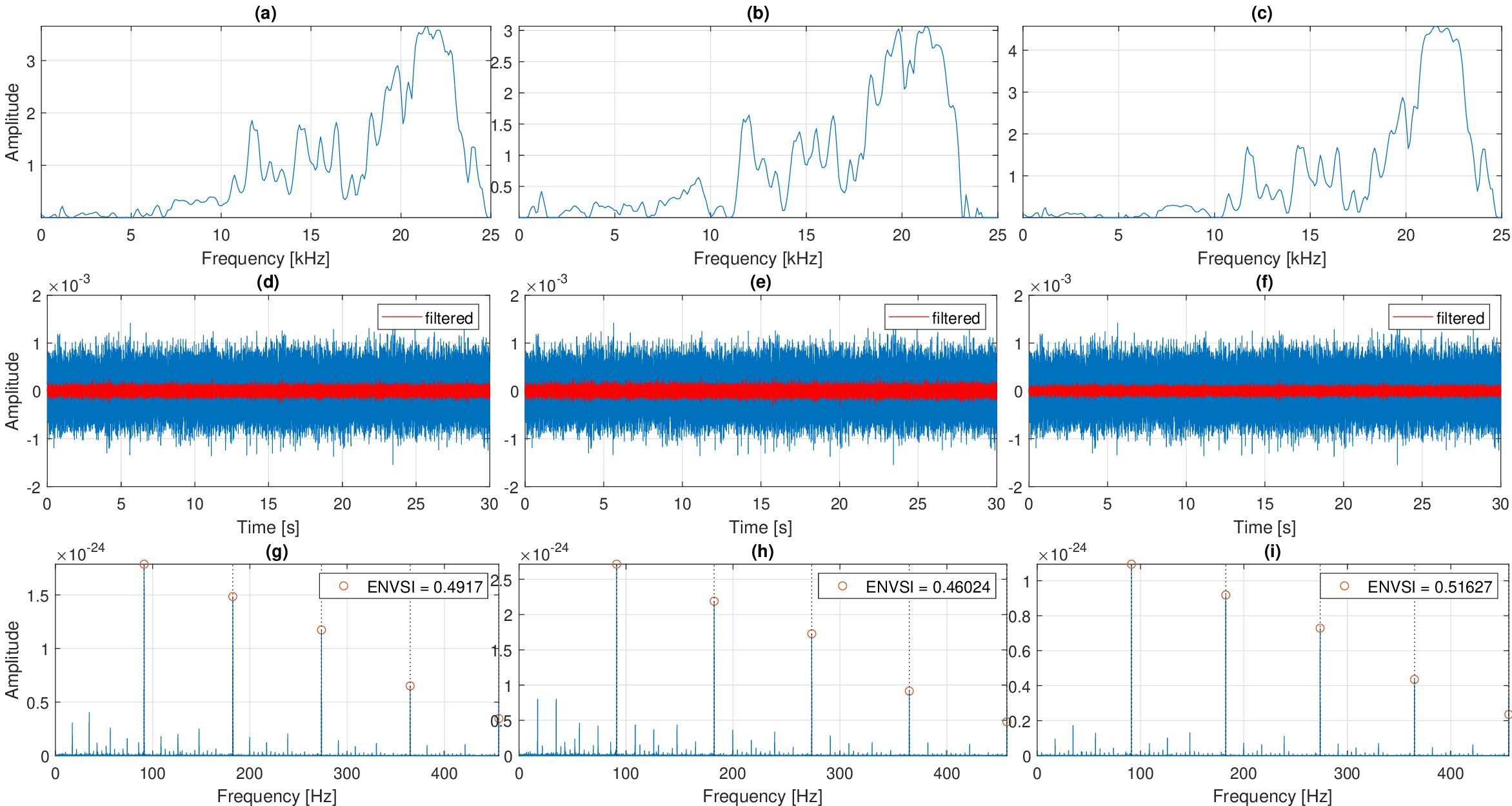}
    \caption{{Selectors results for Case 2: (a) NTF $\beta=-1$ (b) NTF $\beta=0$ (c) NTF $\beta=1$ for the real vibration signal, filtration results (d) - (f), and SES of the filtered signals with  given selectors (g) - (i), respectively.}}
    \label{fig:final_NTF_real_case_2}
\end{figure}

{In Fig. \ref{fig:final_NTF_real_case_2} the final results for the presented approach are presented. The selector characteristics, see subplots (a) - (c), are presented in the same convention for 3 considered values of $\beta$. In the presented case, the fault component and their harmonics can be observed for all considered $\beta$. The ENVSI value for $\beta=-1$ is eq, which means that the extracted information is clear. }

\subsubsection{Comparison with popular selectors}\label{com_real}
\begin{figure} [h!]
    \centering
    \includegraphics[width=0.8\linewidth]{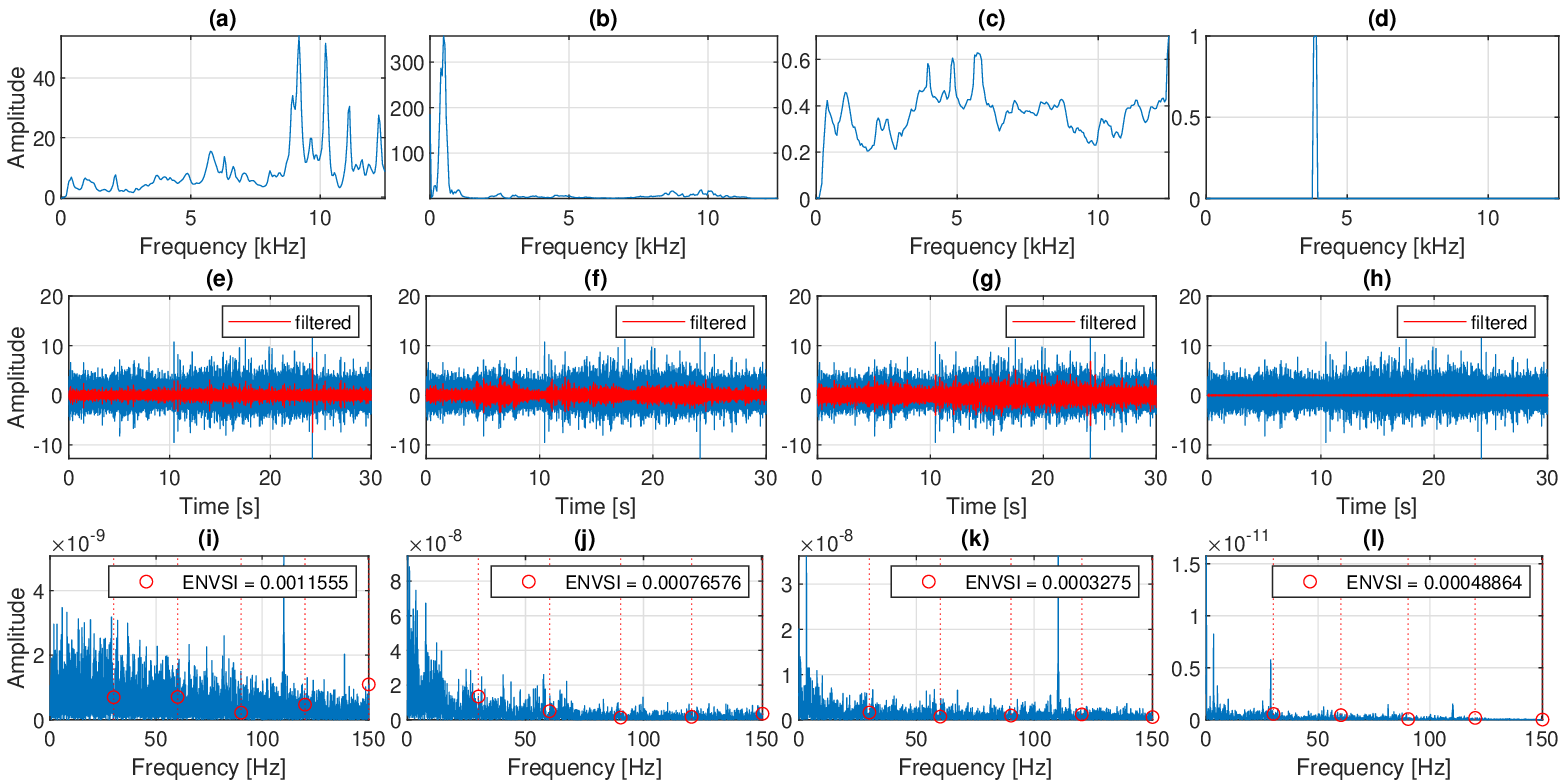}
    \caption{{Selectors results for Case 1: (a) spectral kurtosis, (b) CV, (c) alpha (d) Pearson for the real vibration signal, filtration results (e) - (h), and SES of the filtered signals with given selectors (i) - (l), respectively.}}
    \label{fig:comparison_real}
\end{figure}

\begin{figure} [h!]
    \centering
    \includegraphics[width=0.8\linewidth]{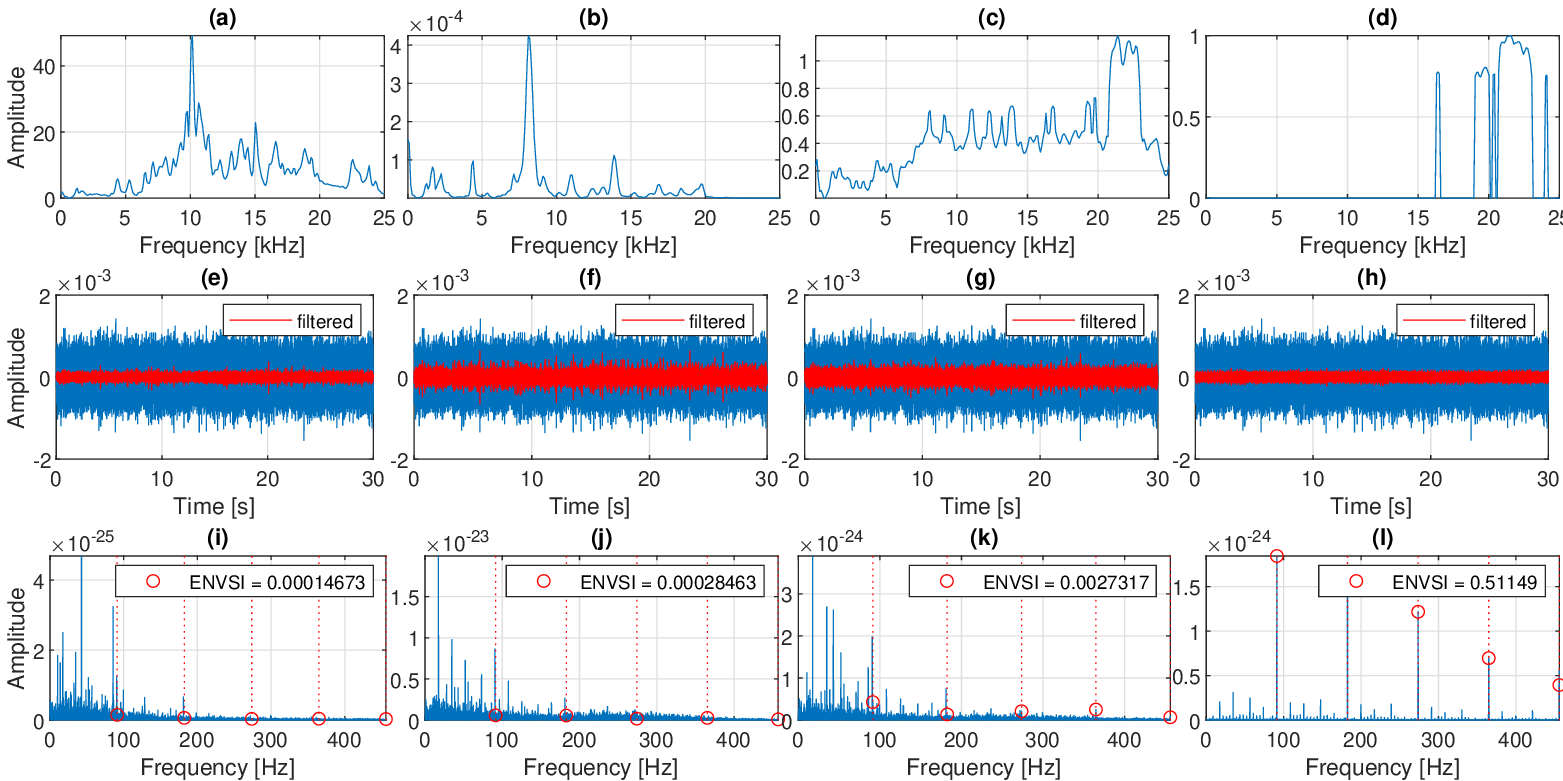}
    \caption{{Selectors results for Case 2: (a) spectral kurtosis, (b) CV, (c) alpha (d) Pearson for the real vibration signal, filtration results (e) - (h), and SES of the filtered signals with given selectors (i) - (l), respectively.}}
    \label{fig:comparison_real2}
\end{figure}

The results obtained using our method have been compared with popular selectors that are also based on the spectrogram, namely spectral kurtosis, CV-based selector, Alpha selector, and Pearson-based selector.  The results are presented in Fig.~\ref{fig:comparison_real} and {\ref{fig:comparison_real2} for Case 1 and 2, respectively}. The shape of the selectors, see Fig.~\ref{fig:comparison_real} a-d,  the difference between the raw and filtered signals, see Fig.~\ref{fig:comparison_real} e-h,  and the structure of the squared envelope spectrum after filtration with the given selectors, see Fig.~\ref{fig:comparison_real} i-l, were considered.  {Except spectral kurtosis, which is mentioned here as a classical approach, sensitive to outliers in the signal, the remaining selectors are dedicated to data with impulsive noise. However, none of these selectors provides clear information on the anticipated informative frequency band (2 - 3 kHz). This is due to the small amplitude of SOI, which is hidden in the background noise and makes the detection challenging.}
The second row of pictures, see subplots (e) - (h), shows the original observation (blue) and the signals extracted with the estimated filters. The extracted signals are very weak (see subplot (h)) or pre-processed in a way that the filtered signals do not reveal clear cyclic impulsive behavior. 
Finally, the squared envelope spectrum for the extracted signals is plotted to detect a family of fault frequencies; see subplots (i) - (l).

As can be seen, no fault frequency components are detected and the ENVSI values are very small, so one may conclude that the selection of the informative band is not effective. 
{The results for selectors are presented in Fig. \ref{fig:comparison_real2} for Case 2. The shape of the selectors, as in Case 1, is shown in subplots (a) - (d). The first two selectors, i.e. spectral kurtosis and CV, exclude the information band in total. Only Pearson's selector indicates the IFB correctly. In the second row, see subplots (e) - (h), the raw acoustic data (marked by blue) and signal obtained by the filtration by proposed selectors (marked by red) are presented. Finally, the SES of extracted signals are shown. As can be seen, in the case of the Pearson selector, see subplot (l), the fault component and its harmonics are clearly visible. The ENVSI value, in this case, is equal to 0.51149.  }

{The proposed approach was also compared with the kurtogram, Infogram, and CFFsgram. For Case 1, the results (see Fig. \ref{fig:appendix_Case1_gram} in the Appendix) and the SES of the filtered signals with the ENVSI value (see Fig. \ref{fig:appendix_Case1_SES} in the Appendix) are presented. Similarly, for Case 2, the results (see Fig. \ref{fig:appendix_Case2_gram} in the Appendix) and the SES of the filtered signals with the ENVSI value (see Fig. \ref{fig:appendix_Case2_SES} in the Appendix) are shown.
The aggregate results of the ENVSI values (defined in Eq. (\ref{eq:ENVSI})), sorted in descending order, are illustrated in Fig. \ref{fig:envsi_bar_all}. To enhance visibility, the ENVSI values are presented on a logarithmic scale. Fig. \ref{fig:envsi_bar} provides a comparison for Case 1.

\begin{figure} [h!]
     \centering
     \begin{subfigure}[b]{\textwidth}
         \centering
         \includegraphics[width=0.75\linewidth]{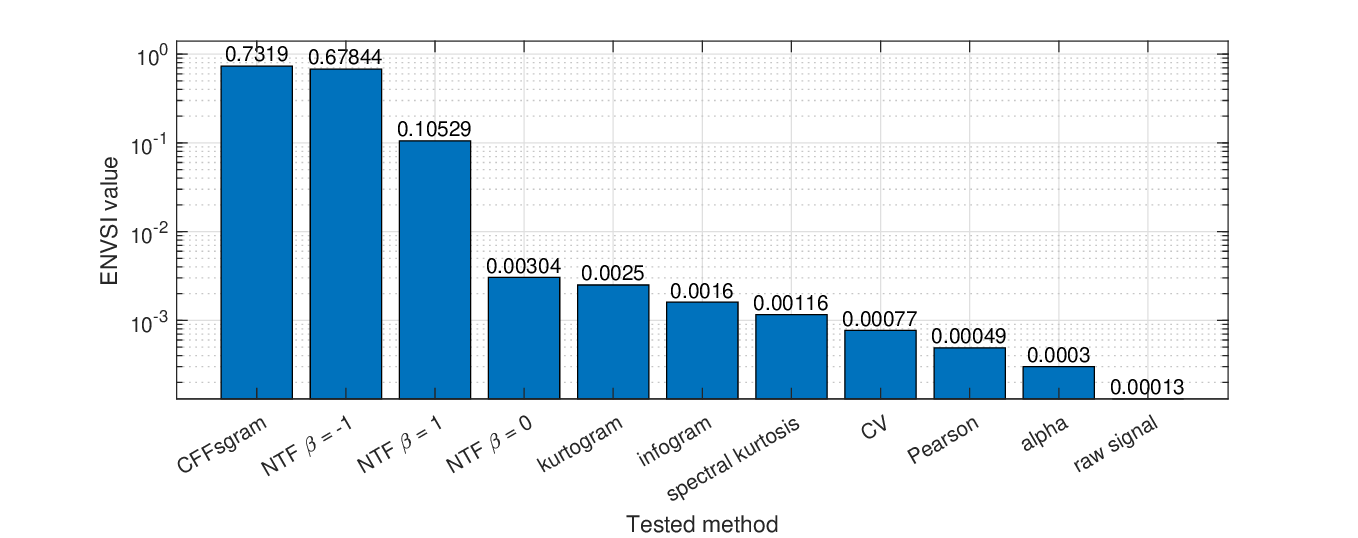}
         \caption{Case 1}
         \label{fig:envsi_bar}
     \end{subfigure}
     \begin{subfigure}[b]{\textwidth}
         \centering
         \includegraphics[width=0.75\linewidth]{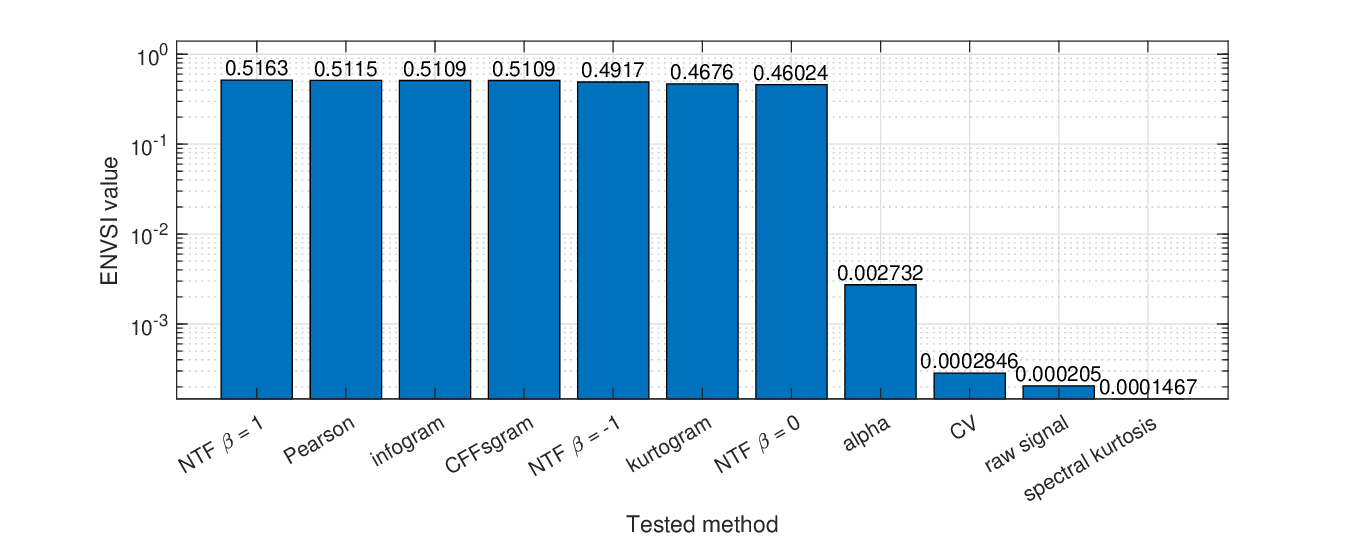}
         \caption{Case 2}
         \label{fig:envsi_bar1}
     \end{subfigure}
        \caption{{Comparison of sorted ENVSI values of the real signals signal, filtered signals with the given selectors and filtered according to the kurtogram, Infogram, and CFFsgram results.}}
        \label{fig:envsi_bar_all}
\end{figure}

As can be seen, for the signal considered with complicated structure selectors based on NTF $\beta=1$ and $\beta=-1$ give effective results and the fault frequency can be easily detected. The CFFsgram achieves the best ENVSI value, equal to 0.7319. However, the proposed methodology does not require prior knowledge of characteristic fault frequency values, whereas the CFFsgram relies on the fault frequency to determine the IFB. Other techniques yield ENVSI values close to 0.001, making the fault frequency and its harmonics indistinguishable from background noise, rendering fault detection impossible. In Case 2, the highest ENVSI values were achieved using the NTF method with $\beta = 1$. Nonetheless, the results are comparable across the different methods. Nevertheless, the results are comparable for most of the compared methods (excluding alpha, CV, and spectral kurtosis, where the ENVSI values are lower than 0.002). }

\section{{Efficiency and computational costs}}

\begin{figure}[h!]
     \centering
     \begin{subfigure}[b]{0.3\textwidth}
         \centering
         \includegraphics[width=\textwidth]{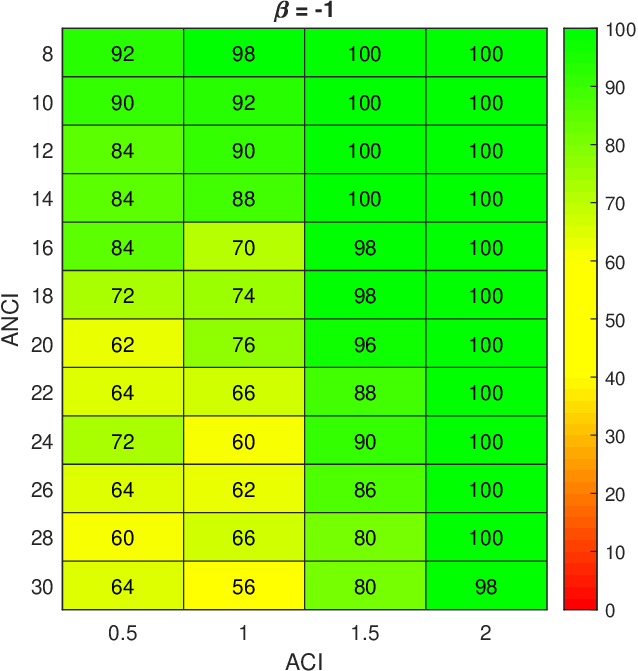}
     \end{subfigure}
        \begin{subfigure}[b]{0.3\textwidth}
         \centering
         \includegraphics[width=\textwidth]{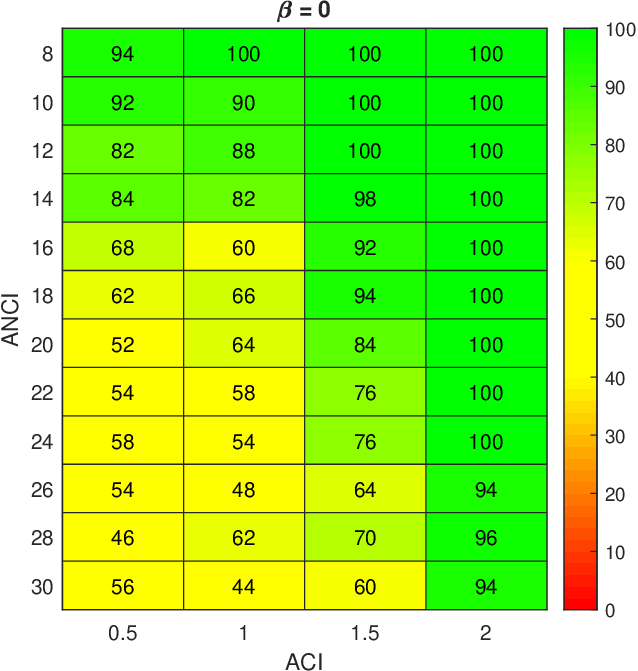}
       
     \end{subfigure}
        \begin{subfigure}[b]{0.3\textwidth}
         \centering
         \includegraphics[width=\textwidth]{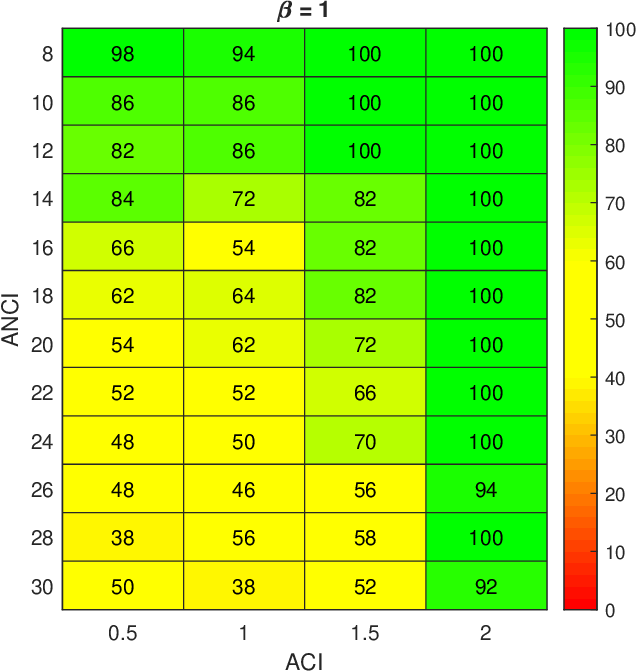}
     \end{subfigure}
        \caption{{Efficiency of the proposed approach for three considered $\beta$ parameters and different values of $A_{CI}$, $A_{NCI}$ based on 50 MC simulations.}}
        \label{fig:efficiency}
\end{figure}

Fig. \ref{fig:efficiency} presents the efficiency (in percent) of the proposed methodology based on the 50 Monte Carlo simulations for different values of $A_{CI}$, $A_{NCI}$ and $\beta = -1$, $ \beta = 0$, and $\beta = 1$, respectively. The parameters of the signals and the number of NTF classes were chosen in the same way as in the simulations, i.e., $M=30$ segments and the number of classes was $K=4$. {The values in the tables represent the percentage of correct results, with red indicating low efficiency, green representing high efficiency, and yellow denoting intermediate performance levels. As can be seen, the best result was received for $\beta = -1$. For the parameter $A_{CI}$ = 2, the efficiency of the proposed method for $\beta = -1$ is above 95\% across all $A_{NCI}$ values. However, as the amplitude of the cyclic pulses decreases and the non-cyclic impulses increase, the efficiency of the method decreases to 56\%.}

 {The efficiency of the proposed approach was also compared with seven benchmark methods, namely, spectral kurtosis, CV, Alpha selector, Pearson selector, kurtogram, Infogram, and CFFsgram. In Figs. \ref{fig:efficiency_sel} and \ref{fig:efficiency_grams} in the Appendix we present their efficiency. In comparison to the proposed method, it is evident that the reference approaches exhibit a much lower degree of efficiency. Methods based on kurtosis, such as spectral kurtosis and kurtogram, can properly detect the fault only where $A_{NCI}$ is at its lowest and $A_{CI}$ is at its highest value. The other three selectors, i.e. CV, Alpha, and Pearson, also provide correct results only for low non-cyclic and high cyclic impulses. Similarly, Infogram works only for the highest $A_{CI}$ values, although it allows slightly higher $A_{NCI}$ values. A noticeable improvement can be observed with the CFFs-gram, where not only the highest $A_{CI}$ is allowed, but also the lowest $A_{NCI}$ values (see Fig. \ref{eff_CFFsgram}). However, it is still significantly less effective than the proposed method. Additionally, the CFFs-gram is not blind. }


\begin{figure}[h!]
     \centering
     \begin{subfigure}[b]{0.49\textwidth}
         \centering
         \includegraphics[width=\textwidth]{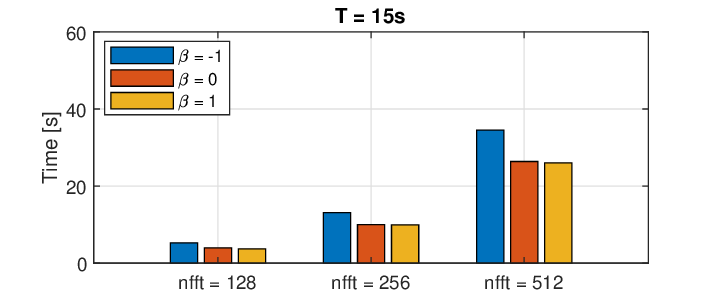}
     \end{subfigure}
     \begin{subfigure}[b]{0.49\textwidth}
         \centering
         \includegraphics[width=\textwidth]{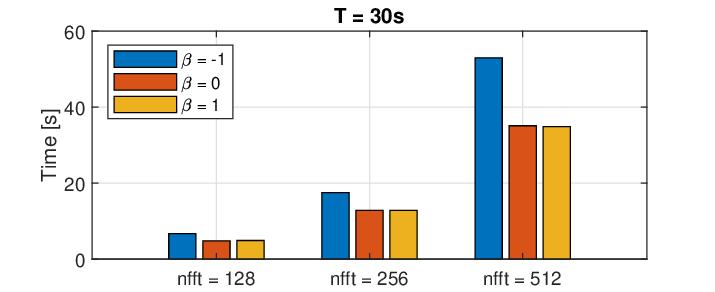}
     \end{subfigure}
        \caption{{Computation time depends on the nfft and the signal length (T) for different values of the parameter $\beta$.}}
        \label{fig:czasy1}
\end{figure}

{Fig. \ref{fig:czasy1} presents the averaged computation time depending on the parameters nfft (number of points in the FFT) and T (signal length in seconds). The calculations were performed in the Matlab environment on a computer with an AMD Ryzen 9 5900X processor and 80 GB of RAM. These parameters directly affect the size of the tensor, which translates into the running time of the NTF algorithm. It can also be seen that the computation time is the longest for $\beta = -1$ in each case. This difference is particularly noticeable with larger nfft values and longer signal time. In comparison, the values of the computation time for $\beta = 0$ and $\beta = 1$ are noticeably lower and similar to each other. However, although the counting time is the longest for the parameter $\beta = -1$, it is only then that the correct results for Case 1 are obtained. }

\section{Conclusions} \label{concl} 
 {The approach outlined in this work provides a novel methodology for detecting local damage in rolling element bearings using vibration signals. The proposed method demonstrates adaptability and potential applicability in various real-world scenarios by addressing the challenges of weak cyclic impulsive signals and strong non-Gaussian noise. Dependence maps and tensor analysis enable identifying the informative frequency band which allows one to obtain SOI and identify the fault frequency in envelope analysis of prefiltered signal.}
    
\textcolor{black}{The proposed method contains original idea of tensor analysis of dependence maps calculated for vibration signal. According to the best knowledge of the authors, the presented approach is novel. Tensor analysis is known and has already been applied to spectrogram representation, however, the proposed approach deals with dependence maps that contain the internal similarity of the signal instead of the energy distribution. The tensor analysis of such representation of data enable to extract very selective frequency band which correspond to local fault.}
       
{The proposed method is fully automatic; nevertheless, the complexity of the signal structure determines the number of classes required for proper decomposition. Based on previous experience, the analyzed case was divided into four classes. The key parameters crucial for the final results are the parameter $\beta$ and the number of classes $K$ as determined by the data analysis.} However, the study reveals that an optimal value of $\beta$ for the real signals analyzed with impulsive noise is $\beta=-1$, which corresponds to the Itakura-Saito distance. It appears that the proposed approach is very effective, and the signal extracted after demodulation provides an excellent structure of the squared envelope spectrum.

This method is intended for situations where a long vibration signal is available and the squared envelope spectrum of the entire signal analysis does not provide clear information about local faults. However, the analysis of shorter fragments of the signal sometimes gives ambiguous information about local faults. This tool uses information from all received signal fragments simultaneously, allowing us to build a filter so that the analysis of the squared envelope spectrum of the filtered signal is easy to interpret.

 The validation process includes both synthetic and real-world vibration {and acoustic} signals, confirming the robustness and effectiveness of the method. The method has been compared with classical and recently developed approaches dedicated to signals with non-Gaussian noise. {One of the compared approach was spectral kurtosis that examines signal impulsivity, one of the main characteristics of SOI. It is a widely used tool in diagnostics. However, it is also known for its sensitivity to large values in the data. Interesting alternatives that also assess signal impulsivity but are less sensitive to large observations are the alpha selector, which is based on the alpha stability index from an alpha-stable distribution, and the CV-based selector, which uses conditional variance. Although all of these methods evaluate whether the SOI exhibits impulsivity, they do not evaluate cyclicity. The cyclicity of the SOI can be tested using the Pearson selector, which is based on the Pearson correlation coefficient. This method is effective in many real-world situations, but it can produce false results in the presence of impulsive noise. The proposed method outperforms the referenced approaches} in the sense of filter selectivity and ENVSI value.  
{ The ENVSI for the real signal from the hammer crusher, which has been filtered using the filter characteristic derived from the NTF with an optimal parameter value of $\beta=-1$, produces a significantly high value of 0.678. The ENVSI values for other methods, which do not require the frequency of damage, chosen for the comparison are smaller than 0.0012, so close to 0. However, the CFFsgram (as a method that requires prior knowledge about the values of those frequencies) achieves a better ENVSI value, equal to 0.7319. This means that for the proposed method there is a high ratio of the energy of the harmonics of the fault frequency to the background of the squared envelope spectrum, making damage detection easy and effective. Thus, compared to other methods, the proposed method (NTF with $\beta=-1$) is better because it is capable of diagnosing damage under difficult conditions. For Case 2, which was measured on the test rig, the results obtained by other methods are comparable in all cases.}

In the future study, several possible improvements may be considered: a more advanced time-frequency representation could be applied, some segment size and number of segments analysis could be performed, different (robust) correlation measures could be tested, and finally, more efficient factorization may be applied. However, the framework of processing will be the same and we do not expect significant improvements (results are already very good).

\section*{Acknowledgments}
The work of RZ and AW is supported by the National Center of Science under the Sheng2 project No. UMO-2021/40/Q/ST8/00024 "NonGauMech - New methods of processing non-stationary signals (identification, segmentation, extraction, modeling) with non-Gaussian characteristics to monitor complex mechanical structures".

\bibliography{bibfile}

@article{berry2007algorithms,
  title={Algorithms and applications for approximate nonnegative matrix factorization},
  author={Berry, Michael W and Browne, Murray and Langville, Amy N and Pauca, V Paul and Plemmons, Robert J},
  journal={Computational statistics \& data analysis},
  volume={52},
  number={1},
  pages={155--173},
  year={2007},
  publisher={Elsevier}
}

@article{sra2005generalized,
  title={Generalized nonnegative matrix approximations with Bregman divergences},
  author={Sra, Suvrit and Dhillon, Inderjit},
  journal={Advances in neural information processing systems},
  volume={18},
  year={2005}
}

@article{pit2016,
  title={THE 20-60-20 RULE.},
  author={Jaworski, Piotr and Pitera, Marcin},
  journal={Discrete \& Continuous Dynamical Systems-Series B},
  volume={21},
  number={4},
  year={2016}
}

@article{pit2019,
  title={New fat-tail normality test based on conditional second moments with applications to finance},
  author={Jelito, Damian and Pitera, Marcin},
  journal={Available at SSRN 3303132},
  year={2018}
}

@article{Hebda-Sobkowicz2020mssp,
author={Hebda-Sobkowicz, Justyna and Zimroz, Rados{\l}aw and Pitera, Marcin and Wyłomańska, Agnieszka},
title={Informative frequency band selection in the presence of non-Gaussian noise – a novel approach based on the conditional variance statistic with application to bearing fault diagnosis},
journal={Mechanical Systems and Signal Processing},
year={2020},
volume={145},
pages={106971},
}

@article{cul,
  title={Simple consistent estimators of stable distribution parameters},
  author={McCulloch, J Huston},
  journal={Communications in Statistics-Simulation and Computation},
  volume={15},
  number={4},
  pages={1109--1136},
  year={1986},
  publisher={Taylor \& Francis}
}

@book{Taqqu,
  title={Stable Non-Gaussian Random Processes: Stochastic Models with Infinite Variance},
  author={Samoradnitsky, Gennady},
  year={1994},
  publisher={Chapman \& Hall}
}

@article{pachaud1997crest,
  title={Crest factor and kurtosis contributions to identify defects inducing periodical impulsive forces},
  author={Pachaud, C and Salvetat, R and Fray, C},
  journal={Mechanical Systems and Signal Processing},
  volume={11},
  number={6},
  pages={903--916},
  year={1997},
  publisher={Elsevier}
}

@article{BARSZCZ2011431,
title = "A novel method for the optimal band selection for vibration signal demodulation and comparison with the Kurtogram",
journal = "Mechanical Systems and Signal Processing",
volume = "25",
number = "1",
pages = "431 - 451",
year = "2011",
issn = "0888-3270",
author = "Tomasz Barszcz and Adam Jab{\l}o{\'n}ski",
}

@article{feng2013recent,
  title={Recent advances in time--frequency analysis methods for machinery fault diagnosis: a review with application examples},
  author={Feng, Zhipeng and Liang, Ming and Chu, Fulei},
  journal={Mechanical Systems and Signal Processing},
  volume={38},
  number={1},
  pages={165--205},
  year={2013},
  publisher={Elsevier}
}

@article{nowicki2021dependency,
author = {Jakub Nowicki and Justyna Hebda-Sobkowicz and Radosław Zimroz and Agnieszka Wyłomańska},
title = {Dependency measures for the diagnosis of local faults in application to the heavy-tailed vibration signal},
journal = {Applied Acoustics},
volume = {178},
pages = {107974},
year = {2021},
issn = {0003-682X},
}

@book{dunn2009basic,
  title={Basic statistics: a primer for the biomedical sciences},
  author={Dunn, Olive Jean and Clark, Virginia A},
  year={2009},
  publisher={John Wiley \& Sons}
}

@article{emp_kurt,
  title={Comparing measures of sample skewness and kurtosis},
  author={Joanes, DN and Gill, CA},
  journal={Journal of the Royal Statistical Society: Series D (The Statistician)},
  volume={47},
  number={1},
  pages={183--189},
  year={1998},
  publisher={Wiley Online Library}
}

@book{boash,
  title={Time-frequency signal analysis and processing: a comprehensive reference},
  author={Boashash, Boualem},
  year={2015},
  publisher={Academic press}
}

@article{kurt,
  title={Kurtosis as peakedness, 1905--2014. R.I.P.},
  author={Westfall, Peter H},
  journal={The American Statistician},
  volume={68},
  number={3},
  pages={191--195},
  year={2014},
  publisher={Taylor \& Francis}
}

@article{wodecki2021local,
  title={Local damage detection based on vibration data analysis in the presence of Gaussian and heavy-tailed impulsive noise},
  author={Wodecki, Jacek and Michalak, Anna and Zimroz, Rados{\l}aw},
  journal={Measurement},
  volume={169},
  pages={108400},
  year={2021},
  publisher={Elsevier}
}

@ARTICLE{Gabor20242944,
	author = {Gabor, Mateusz and Zdunek, Rafal and Zimroz, Radoslaw and Wylomanska, Agnieszka},
	title = {Bearing Damage Detection With Orthogonal and Nonnegative Low-Rank Feature Extraction},
	year = {2024},
	journal = {IEEE Transactions on Industrial Informatics},
	volume = {20},
	number = {2},
	pages = {2944 – 2955}
	}

@ARTICLE{Gabor2023,
	author = {Gabor, Mateusz and Zdunek, Rafal and Zimroz, Radoslaw and Wodecki, Jacek and Wylomanska, Agnieszka},
	title = {Non-negative tensor factorization for vibration-based local damage detection},
	year = {2023},
	journal = {Mechanical Systems and Signal Processing},
	volume = {198}
	}

@article{kruczek2020detect,
  title={How to detect the cyclostationarity in heavy-tailed distributed signals},
  author={Kruczek, Piotr and Zimroz, Rados{\l}aw and Wy{\l}oma{\'n}ska, Agnieszka},
  journal={Signal Processing},
  volume={172},
  pages={107514},
  year={2020},
  publisher={Elsevier}
}

@article{wylomanska2016impulsive,
  title={Impulsive noise cancellation method for copper ore crusher vibration signals enhancement},
  author={Wy{\l}oma{\'n}ska, Agnieszka and Zimroz, Rados{\l}aw and Janczura, Joanna and Obuchowski, Jakub},
  journal={IEEE Transactions on Industrial Electronics},
  volume={63},
  number={9},
  pages={5612--5621},
  year={2016},
  publisher={IEEE}
}

@article{Mauricio2020,
author={Mauricio, A. and Qi, J. and Smith, W.A. and Sarazin, M. and Randall, R.B. and Janssens, K. and Gryllias, K.},
title={{Bearing diagnostics under strong electromagnetic interference based on Integrated Spectral Coherence}},
journal={Mechanical Systems and Signal Processing},
year={2020},
volume={140},
pages={106673},
}

@article{borghesani2017cs2,
  title={{CS2 analysis in presence of non-Gaussian background noise -- Effect on traditional estimators and resilience of log-envelope indicators}},
  author={Borghesani, P and Antoni, J},
  journal={Mechanical Systems and Signal Processing},
  volume={90},
  pages={378--398},
  year={2017}
}

@ARTICLE{Yu2013155,
	author = {Yu, Gang and Li, Changning and Zhang, Jianfeng},
	title = {{A new statistical modeling and detection method for rolling element bearing faults based on alpha-stable distribution}},
	year = {2013},
	journal = {Mechanical Systems and Signal Processing},
	volume = {41},
	number = {1-2},
	pages = {155 – 175}
}

@article{miao2017improvement, 
  title={Improvement of kurtosis-guided-grams via Gini index for bearing fault feature identification},
  author={Miao, Yonghao and Zhao, Ming and Lin, Jing},
  journal={Measurement Science and Technology},
  volume={28},
  number={12},
  pages={125001},
  year={2017},
  publisher={IOP Publishing}
}

@article{wang2013enhanced,
  title={{An enhanced Kurtogram method for fault diagnosis of rolling element bearings}},
  author={Wang, Dong and Peter, W Tse and Tsui, Kwok Leung},
  journal={Mechanical Systems and Signal Processing},
  volume={35},
  number={1-2},
  pages={176--199},
  year={2013},
  publisher={Elsevier}
}

@article{peter2013design,
  title={{The design of a new sparsogram for fast bearing fault diagnosis: Part 1 of the two related manuscripts that have a joint title as “Two automatic vibration-based fault diagnostic methods using the novel sparsity measurement--Parts 1 and 2”}},
  author={Peter, W Tse and Wang, Dong},
  journal={Mechanical Systems and Signal Processing},
  volume={40},
  number={2},
  pages={499--519},
  year={2013},
  publisher={Elsevier}
}

@article{liu2019accugram,
  title={ACCUGRAM: A novel approach based on classification to frequency band selection for rotating machinery fault diagnosis},
  author={Liu, Zhiliang and Jin, Yaqiang and Zuo, Ming J and Peng, Dandan},
  journal={ISA Transactions},
  volume={95},
  pages={346--357},
  year={2019},
  publisher={Elsevier}
}

@article{mauricio2020improved,
  title={Improved Envelope Spectrum via Feature Optimisation-gram (IESFOgram): A novel tool for rolling element bearing diagnostics under non-stationary operating conditions},
  author={Mauricio, Alexandre and Smith, Wade A and Randall, Robert B and Antoni, Jerome and Gryllias, Konstantinos},
  journal={Mechanical Systems and Signal Processing},
  volume={144},
  pages={106891},
  year={2020},
  publisher={Elsevier}
}

@article{moshrefzadeh2018autogram,
  title={{The Autogram: An effective approach for selecting the optimal demodulation band in rolling element bearings diagnosis}},
  author={Moshrefzadeh, Ali and Fasana, Alessandro},
  journal={Mechanical Systems and Signal Processing},
  volume={105},
  pages={294--318},
  year={2018},
  publisher={Elsevier}
}

@article{wu2021enkurgram,
  title={{The Enkurgram: A characteristic frequency extraction method for fluid machinery based on multi-band demodulation strategy}},
  author={Wu, Kelin and Chu, Ning and Wu, Dazhuan and Antoni, J{\'e}r{\^o}me},
  journal={Mechanical Systems and Signal Processing},
  volume={155},
  pages={107564},
  year={2021},
  publisher={Elsevier}
}

@article{wang2016extension,
  title={{An extension of the infograms to novel Bayesian inference for bearing fault feature identification}},
  author={Wang, Dong},
  journal={Mechanical Systems and Signal Processing},
  volume={80},
  pages={19--30},
  year={2016},
  publisher={Elsevier}
}

@ARTICLE{Kolda08,
  author = {T. G. Kolda and B. W. Bader},
  title = {Tensor Decompositions and Applications},
  journal = {SIAM Review},
  year = {2009},
  volume = {51},
  pages = {455-500},
  number = {3}
}

@Article{app9183642,
AUTHOR = {Liang, Lin and Wen, Haobin and Liu, Fei and Li, Guang and Li, Maolin},
TITLE = {Feature Extraction of Impulse Faults for Vibration Signals Based on Sparse Non-Negative Tensor Factorization},
JOURNAL = {Applied Sciences},
VOLUME = {9},
YEAR = {2019},
NUMBER = {18},
ARTICLE-NUMBER = {3642},
}

@INPROCEEDINGS{Shashua2005,
  author = {A. Shashua and T. Hazan},
  title = {Non-negative tensor factorization with applications to statistics
    and computer vision},
  booktitle = {Proc. of the 22-th International Conference on Machine Learning},
  year = {2005},
  address = {Bonn, Germany}
}

@incollection{carroll1989fitting,
  title={Fitting of the latent class model via iteratively reweighted least squares CANDECOMP with nonnegativity constraints},
  author={Carroll, J Douglas and De Soete, Geert and Pruzansky, Sandra},
  booktitle={Multiway data analysis},
  pages={463--472},
  year={1989}
}

@ARTICLE{6588559,
  author={Figueiredo, Marisa and Ribeiro, Bernardete and de Almeida, Ana},
  journal={IEEE Transactions on Instrumentation and Measurement}, 
  title={Electrical Signal Source Separation Via Nonnegative Tensor Factorization Using On Site Measurements in a Smart Home}, 
  year={2014},
  volume={63},
  number={2},
  pages={364-373},
  }

@ARTICLE{8497054,
  author={Xiong, Fengchao and Qian, Yuntao and Zhou, Jun and Tang, Yuan Yan},
  journal={IEEE Transactions on Geoscience and Remote Sensing}, 
  title={Hyperspectral Unmixing via Total Variation Regularized Nonnegative Tensor Factorization}, 
  year={2019},
  volume={57},
  number={4},
  pages={2341-2357}
  }

@ARTICLE{FitzGerald2008,
  author = {D. FitzGerald and M. Cranitch and E. Coyle},
  title = {Extended Nonnegative Tensor Factorisation Models for Musical Sound
    Source Separation},
  journal = {Comput Intell Neurosci},
  year = {2008},
  volume = {872425},
  pages = {1--15},
  __markedentry = {[Rafal:1]},
  owner = {HuyAnh},
  timestamp = {2009.01.20}
}

@article{FabienM,
  author    = {Fabien Millioz and
               Nadine Martin},
  title     = {Circularity of the {STFT} and Spectral Kurtosis for Time-Frequency Segmentation in Gaussian Environment},
  journal   = {{IEEE} Trans. Signal Process.},
  volume    = {59},
  number    = {2},
  pages     = {515--524},
  year      = {2011},
  
}

@ARTICLE{4480140,
  author={Huillery, Julien and Millioz, Fabien and Martin, Nadine},
  journal={IEEE Transactions on Signal Processing}, 
  title={On the Description of Spectrogram Probabilities With a Chi-Squared Law}, 
  year={2008},
  volume={56},
  number={6},
  pages={2249-2258}
}

@article{Wodecki2020,
  title={Separation of multiple local-damage-related components from vibration data using Nonnegative Matrix Factorization and multichannel data fusion},
  author={Wodecki, Jacek and Michalak, Anna and Zimroz, Rados{\l}aw and Wy{\l}oma{\'n}ska, Agnieszka},
  journal={Mechanical Systems and Signal Processing},
  volume={145},
  pages={106954},
  year={2020},
  publisher={Elsevier}
}

@article{wodecki2019impulsive,
title={Impulsive source separation using combination of Nonnegative Matrix Factorization of bi-frequency map, spatial denoising and Monte Carlo simulation},
author={Wodecki, Jacek  and Michalak, Anna and Zimroz, Rados{\l}aw and Barszcz, Tomasz and Wy{\l}oma{\'n}ska, Agnieszka},
journal={Mechanical Systems and Signal Processing},
pages={89--101},
year={2019},
volume={127},
}

@article{wang2016new,
  title={A new SKRgram based demodulation technique for planet bearing fault detection},
  author={Wang, Tianyang and Han, Qinkai and Chu, Fulei and Feng, Zhipeng},
  journal={Journal of Sound and Vibration},
  volume={385},
  pages={330--349},
  year={2016},
  publisher={Elsevier}
}

@article{hebda2022infogram,
  title={Infogram performance analysis and its enhancement for bearings diagnostics in presence of non-Gaussian noise},
  author={Hebda-Sobkowicz, Justyna and Zimroz, Rados{\l}aw and Wy{\l}oma{\'n}ska, Agnieszka and Antoni, Jerome},
  journal={Mechanical Systems and Signal Processing},
  volume={170},
  pages={108764},
  year={2022},
  publisher={Elsevier}
}

@article{HebdaSobkowicz2020,
  year = {2020},
  month = apr,
  publisher = {{MDPI} {AG}},
  volume = {10},
  number = {8},
  pages = {2657},
  author = {Justyna Hebda-Sobkowicz and Rados{\l}aw Zimroz and Agnieszka Wy{\l}oma{\'{n}}ska},
  title = {Selection of the Informative Frequency Band in a Bearing Fault Diagnosis in the Presence of Non-{G}aussian Noise -- Comparison of Recently Developed Methods},
  journal = {Applied Sciences}
}

@article{antoni2016info,
  title={The infogram: Entropic evidence of the signature of repetitive transients},
  author={Antoni, Jerome},
  journal={Mechanical Systems and Signal Processing},
  volume={74},
  pages={73--94},
  year={2016},
  publisher={Elsevier}
  }

@article{antoni2006spectral,
  title={The spectral kurtosis: application to the vibratory surveillance and diagnostics of rotating machines},
  author={Antoni, J{\'e}r{\^o}me and Randall, RB},
  journal={Mechanical Systems and Signal Processing},
  volume={20},
  number={2},
  pages={308--331},
  year={2006},
  publisher={Elsevier}
}

@article{antoni2007fast,
  title={Fast computation of the kurtogram for the detection of transient faults},
  author={Antoni, Jerome},
  journal={Mechanical Systems and Signal Processing},
  volume={21},
  number={1},
  pages={108--124},
  year={2007},
  publisher={Elsevier}
}

@article{wodecki2018optimal,
title = {Optimal filter design with progressive genetic algorithm for local damage detection in rolling bearings},
author = {Jacek Wodecki and Anna Michalak and Radoslaw Zimroz},
year = {2018},
date = {2018-01-01},
journal = {Mechanical Systems and Signal Processing},
volume = {102},
pages = {102--116},
publisher = {Elsevier}
}

@article{randall2011rolling,
  title={Rolling element bearing diagnostics - a tutorial},
  author={Randall, Robert B and Antoni, Jerome},
  journal={Mechanical Systems and Signal Processing},
  volume={25},
  number={2},
  pages={485--520},
  year={2011},
  publisher={Elsevier}
}

@book{cichocki2009nonnegative,
  title={Nonnegative matrix and tensor factorizations: applications to exploratory multi-way data analysis and blind source separation},
  author={Cichocki, Andrzej and Zdunek, Rafal and Phan, Anh Huy and Amari, Shun-ichi},
  year={2009},
  publisher={John Wiley \& Sons}
}

@article{allen1977short,
  title={Short term spectral analysis, synthesis, and modification by discrete {F}ourier transform},
  author={Allen, Jonathan},
  journal={IEEE Transactions on Acoustics, Speech, and Signal Processing},
  volume={25},
  number={3},
  pages={235--238},
  year={1977},
  publisher={IEEE}
}

@article{yang2024review,
  title={A review on the application of blind source separation in vibration analysis of mechanical systems},
  author={Yang, Yunxi and Xie, Ruili and Li, Ming and Cheng, Wei},
  journal={Measurement},
  pages={114241},
  year={2024},
  publisher={Elsevier}
}

@article{zhou2023cffsgram,
  title={CFFsgram: A candidate fault frequencies-based optimal demodulation band selection method for axle-box bearing fault diagnosis},
  author={Zhou, Ning and Cheng, Yao and Wang, Zhiwei and Chen, Bingyan and Zhang, Weihua},
  journal={Measurement},
  volume={207},
  pages={112368},
  year={2023},
  publisher={Elsevier}
}

@article{wang2015non,
  title={Non-negative EMD manifold for feature extraction in machinery fault diagnosis},
  author={Wang, Cong and Gan, Meng and others},
  journal={Measurement},
  volume={70},
  pages={188--202},
  year={2015},
  publisher={Elsevier}
}

@article{lee1999learning,
  title={Learning the parts of objects by non-negative matrix factorization},
  author={Lee, Daniel D and Seung, H Sebastian},
  journal={nature},
  volume={401},
  number={6755},
  pages={788--791},
  year={1999},
  publisher={Nature Publishing Group UK London}
}

@article{wang2021intelligent,
  title={Intelligent fault diagnosis of diesel engine via adaptive VMD-Rihaczek distribution and graph regularized bi-directional NMF},
  author={Wang, Xu and Cai, Yanping and Li, Aihua and Zhang, Wei and Yue, Yingjuan and Ming, Anbo},
  journal={Measurement},
  volume={172},
  pages={108823},
  year={2021},
  publisher={Elsevier}
}

@article{liang2016feature,
  title={Feature selection for machine fault diagnosis using clustering of non-negation matrix factorization},
  author={Liang, Lin and Liu, Fei and Li, Maolin and He, Kangkang and Xu, Guanghua},
  journal={Measurement},
  volume={94},
  pages={295--305},
  year={2016},
  publisher={Elsevier}
}

@article{cui2024spectral,
  title={A spectral coherence cyclic periodic index optimization-gram for bearing fault diagnosis},
  author={Cui, Lingli and Zhao, Xinyuan and Liu, Dongdong and Wang, Huaqing},
  journal={Measurement},
  volume={224},
  pages={113898},
  year={2024},
  publisher={Elsevier}
}

@article{wodecki2021influence,
  title={Influence of non-Gaussian noise on the effectiveness of cyclostationary analysis--Simulations and real data analysis},
  author={Wodecki, Jacek and Michalak, Anna and Wy{\l}oma{\'n}ska, Agnieszka and Zimroz, Rados{\l}aw},
  journal={Measurement},
  volume={171},
  pages={108814},
  year={2021},
  publisher={Elsevier}
}

@ARTICLE{Zhao2024335,
	author = {Zhao, Dezun and Shao, Depei and Cui, Lingli},
	title = {CTNet: A data-driven time-frequency technique for wind turbines fault diagnosis under time-varying speeds},
	year = {2024},
	journal = {ISA Transactions},
	volume = {154},
	pages = {335 – 351}
}

@article{ZHAO2024111112,
title = {Frequency-chirprate synchrosqueezing-based scaling chirplet transform for wind turbine nonstationary fault feature time–frequency representation},
journal = {Mechanical Systems and Signal Processing},
volume = {209},
pages = {111112},
year = {2024},
issn = {0888-3270},
author = {Dezun Zhao and Honghao Wang and Lingli Cui},
}

@article{zheng2024progressive,
  title={A progressive multi-source domain adaptation method for bearing fault diagnosis},
  author={Zheng, Xiaorong and He, Zhiwei and Nie, Jiahao and Li, Ping and Dong, Zhekang and Gao, Mingyu},
  journal={Applied Acoustics},
  volume={216},
  pages={109797},
  year={2024},
  publisher={Elsevier}
}

@article{ye2023intelligent,
  title={Intelligent fault diagnosis of rolling bearing using variational mode extraction and improved one-dimensional convolutional neural network},
  author={Ye, Maoyou and Yan, Xiaoan and Chen, Ning and Jia, Minping},
  journal={Applied Acoustics},
  volume={202},
  pages={109143},
  year={2023},
  publisher={Elsevier}
}

@article{wissbrock2024more,
  title={More than spectrograms: Deep representation learning for machinery fault detection},
  author={Wi{\ss}brock, Peter and Ren, Zhao and Pelkmann, David},
  journal={Applied Acoustics},
  volume={225},
  pages={110178},
  year={2024},
  publisher={Elsevier}
}

@ARTICLE{Luo2024,
	author = {Luo, Jingjie and Shao, Haidong and Lin, Jian and Liu, Bin},
	title = {Meta-learning with elastic prototypical network for fault transfer diagnosis of bearings under unstable speeds},
	year = {2024},
	volume = {245},
		journal = {Reliability Engineering and System Safety},
 pages={110001},
	source = {Scopus}
}

@ARTICLE{Li20227328,
	author = {Li, Xin and Shao, Haidong and Lu, Siliang and Xiang, Jiawei and Cai, Baoping},
	title = {Highly Efficient Fault Diagnosis of Rotating Machinery Under Time-Varying Speeds Using {LSISMM} and Small Infrared Thermal Images},
	year = {2022},
	journal = {IEEE Transactions on Systems, Man, and Cybernetics: Systems},
	volume = {52},
	number = {12},
	pages = {7328 – 7340}
}

@ARTICLE{Han2021,
	author = {Han, Tian and Zhang, Longwen and Yin, Zhongjun and Tan, Andy C.C.},
	title = {Rolling bearing fault diagnosis with combined convolutional neural networks and support vector machine},
	year = {2021},
	journal = {Measurement: Journal of the International Measurement Confederation},
	volume = {177},
pages={109022}
}

@article{cheng2022improved,
  title={An improved envelope spectrum via candidate fault frequency optimization-gram for bearing fault diagnosis},
  author={Cheng, Yao and Wang, Shengbo and Chen, Bingyan and Mei, Guiming and Zhang, Weihua and Peng, Han and Tian, Guangrong},
  journal={Journal of Sound and Vibration},
  volume={523},
  pages={116746},
  year={2022},
  publisher={Elsevier}
}
\section*{Appendix}

\begin{figure} [h!]
     \centering
     \begin{subfigure}[b]{1\textwidth}
         \centering
         \includegraphics[width=1\linewidth]{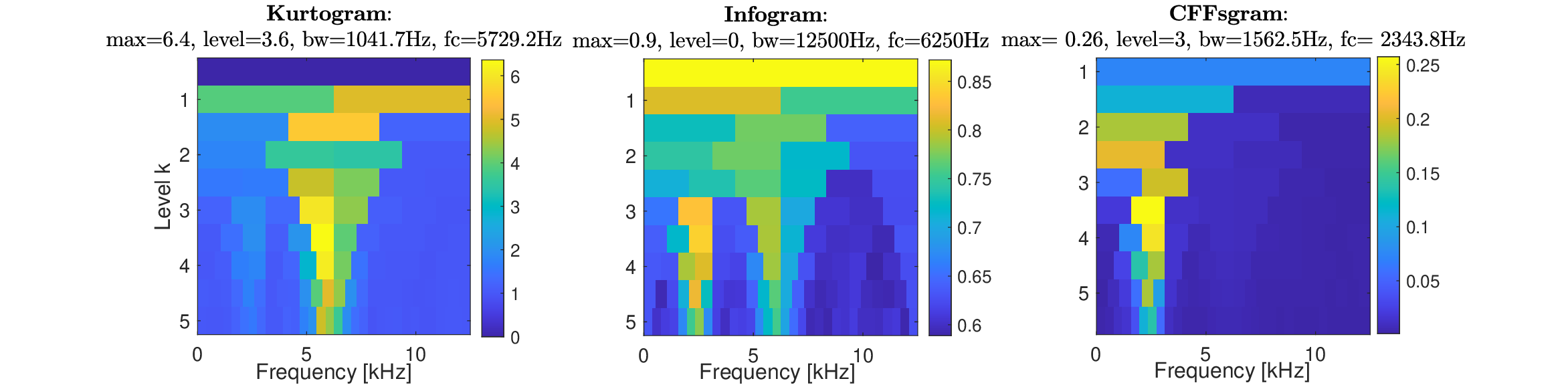}
         \caption{}
         \label{fig:appendix_simul_gram}
     \end{subfigure}
     \begin{subfigure}[b]{\textwidth}
         \centering
         \includegraphics[width=0.8\linewidth]{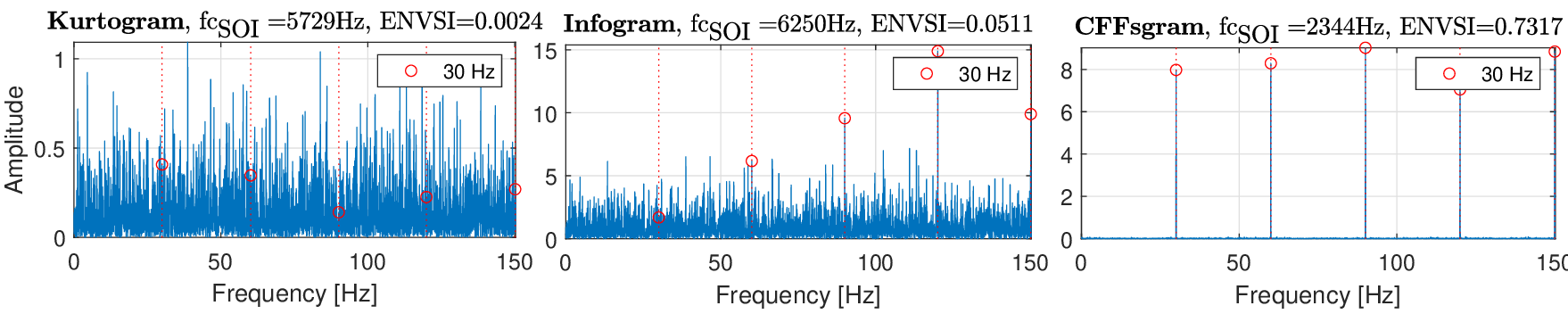}
         \caption{}
         \label{fig:appendix_simul_SES}
     \end{subfigure}
        \caption{{The results (a) and SES of the filtered signals with the ENVSI value (b) of simulated signal analysis by the following methods: kurtogram, infogram, and CFFsgram, respectively.}}
        \label{fig:appendix_simul}
\end{figure}

\begin{figure} [h!]
     \centering
     \begin{subfigure}[b]{1\textwidth}
         \centering
         \includegraphics[width=1\linewidth]{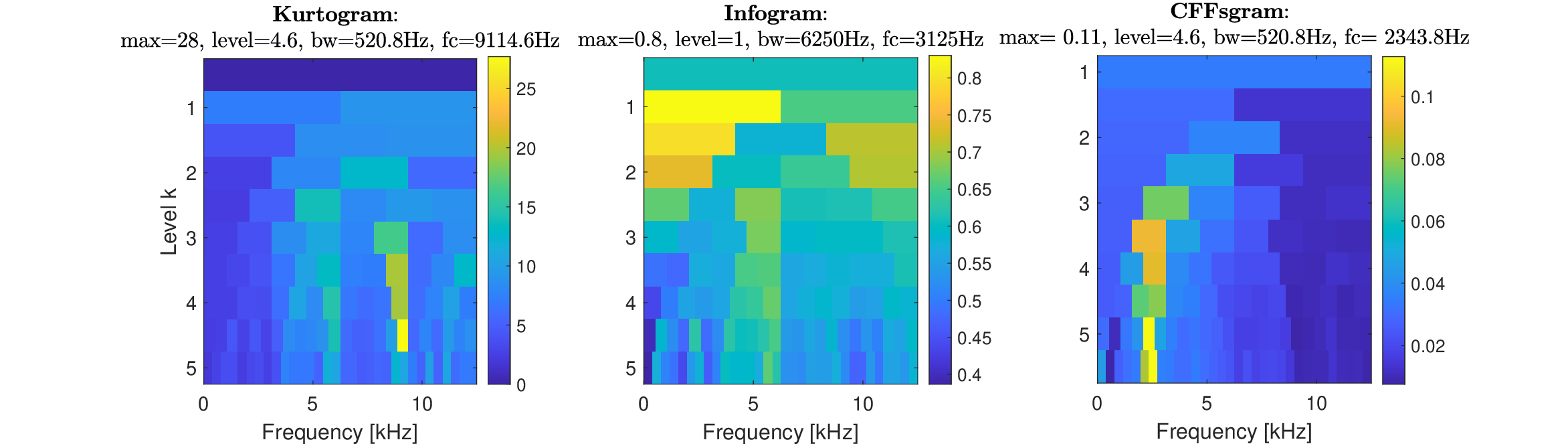}
         \caption{}
         \label{fig:appendix_Case1_gram}
     \end{subfigure}
     \begin{subfigure}[b]{\textwidth}
         \centering
         \includegraphics[width=0.8\linewidth]{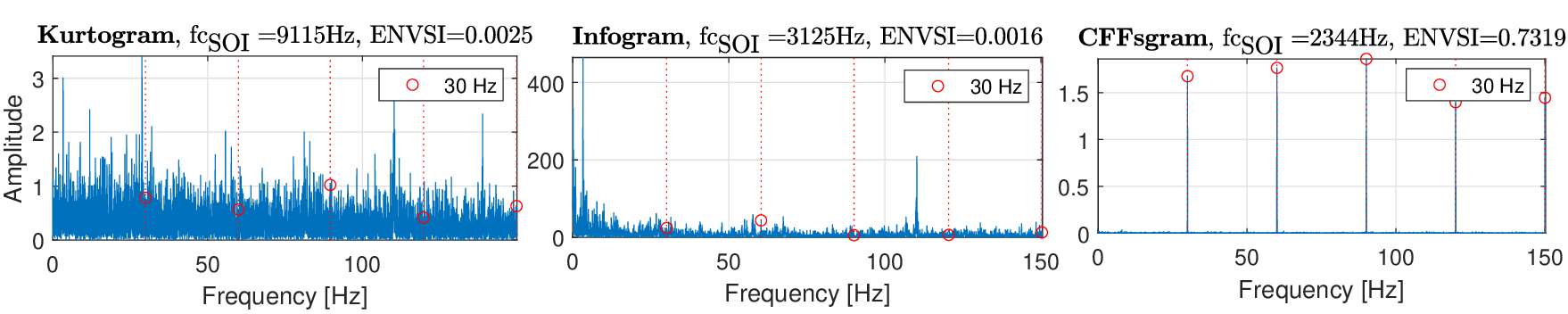}
         \caption{}
         \label{fig:appendix_Case1_SES}
     \end{subfigure}
        \caption{{The results (a) and SES of the filtered signals with the ENVSI value (b) of Case 1 analysis by the following methods: kurtogram, infogram, and CFFsgram, respectively.}}
        \label{fig:appendix_Case1}
\end{figure}

\begin{figure} [ht!]
     \centering
     \begin{subfigure}[b]{1\textwidth}
         \centering
         \includegraphics[width=1\linewidth]{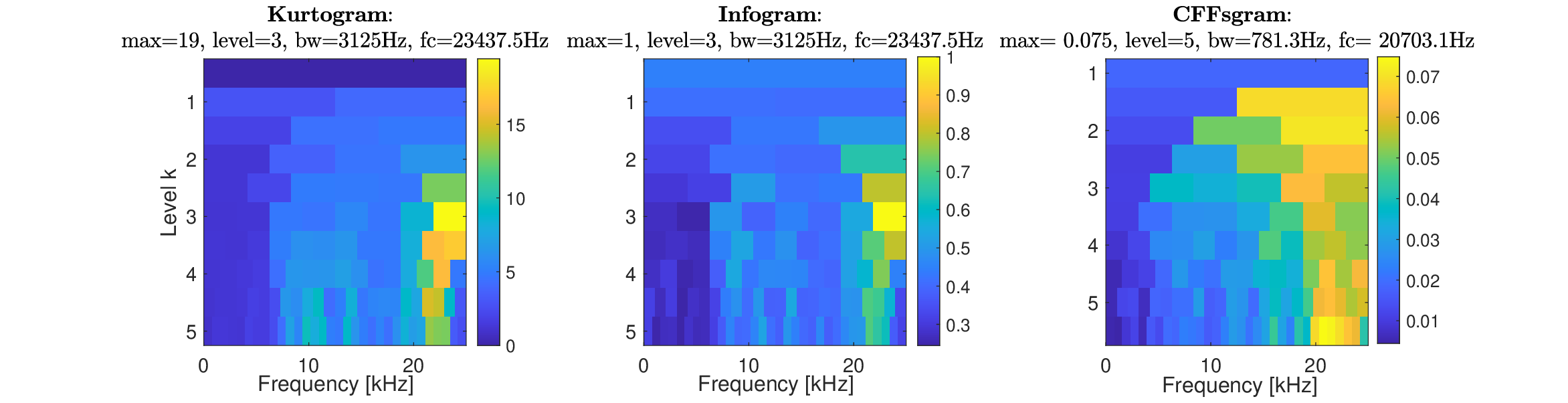}
         \caption{}
         \label{fig:appendix_Case2_gram}
     \end{subfigure}
     \begin{subfigure}[b]{\textwidth}
         \centering
         \includegraphics[width=0.8\linewidth]{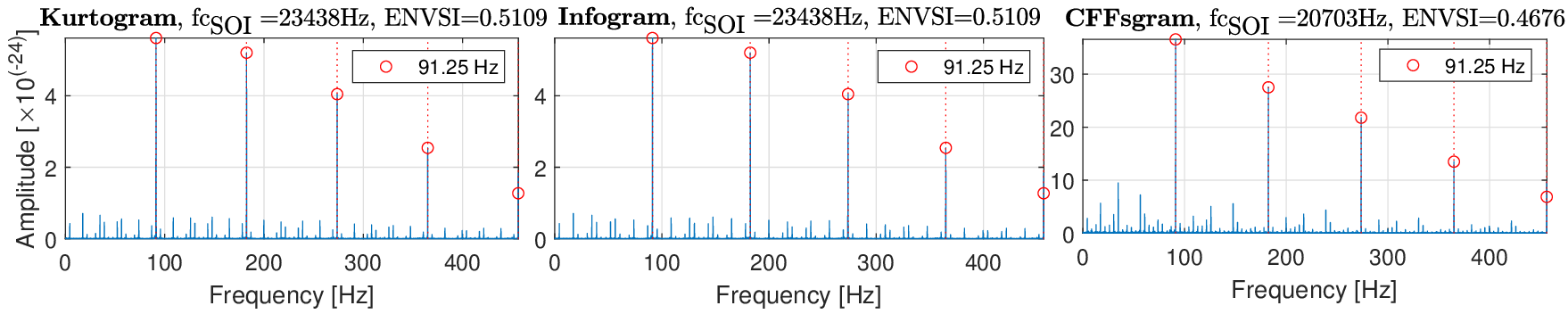}
         \caption{}
         \label{fig:appendix_Case2_SES}
     \end{subfigure}
        \caption{{The results (a) and SES of the filtered signals with the ENVSI value (b) of Case 2 analysis by the following methods: kurtogram, infogram, and CFFsgram, respectively.}}
        \label{fig:appendix_Case2}
\end{figure}

{The comparative results popular and newly proposed methods for all analyzed signals are presented below. 
For the simulated data the results are illustrated in Fig. \ref{fig:appendix_simul}. For the real data namely: vibration signal from hammer crusher (Case 1) and acoustic signal from the test rig (Case 2), are presented in Figs. \ref{fig:appendix_Case1} and \ref{fig:appendix_Case2}, respectively. Detailed descriptions of the methods can be found in the literature, including the kurtogram \cite{antoni2007fast}, Infogram \cite{antoni2016info}, and CFFsgram \cite{zhou2023cffsgram}.}

\begin{figure}[h!]
     \centering
     \begin{subfigure}[b]{0.3\textwidth}
         \centering
         \includegraphics[width=\textwidth]{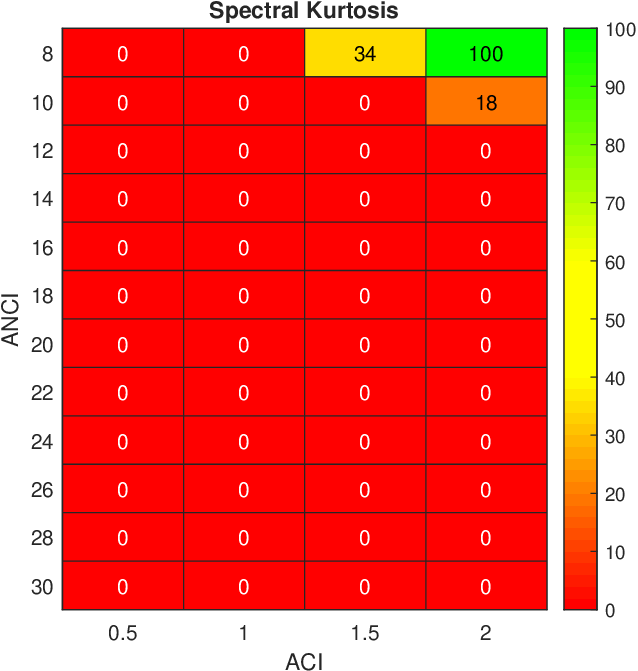}
         \caption{}
         \label{eff_sk}
     \end{subfigure}
        \begin{subfigure}[b]{0.3\textwidth}
         \centering
         \includegraphics[width=\textwidth]{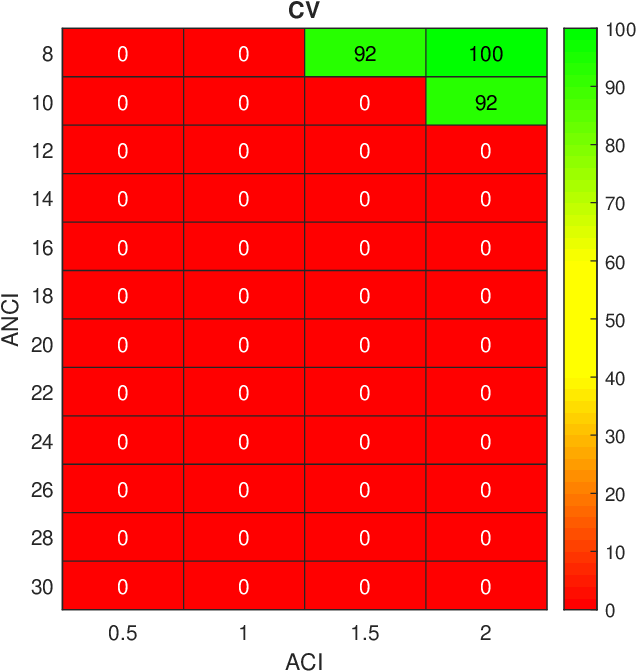}
                  \caption{}
         \label{eff_CV}
     \end{subfigure}
     
        \begin{subfigure}[b]{0.3\textwidth}
         \centering
         \includegraphics[width=\textwidth]{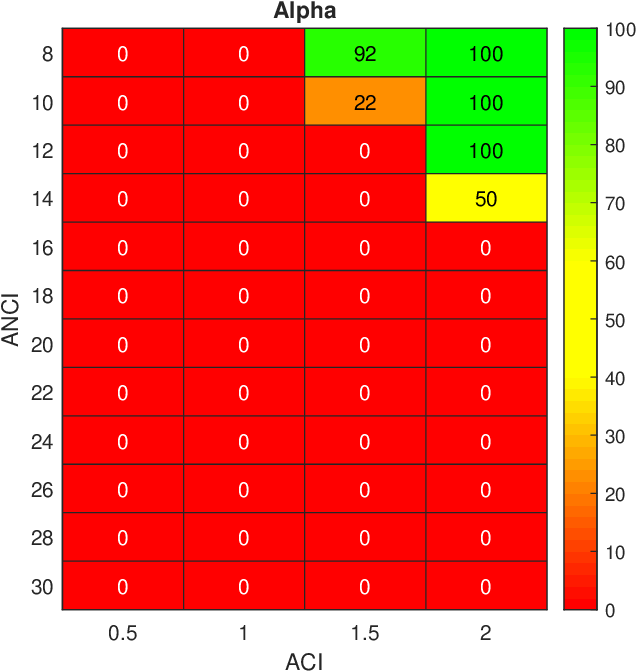}
                  \caption{}
         \label{eff_alpha}
     \end{subfigure}
        \begin{subfigure}[b]{0.3\textwidth}
         \centering
         \includegraphics[width=\textwidth]{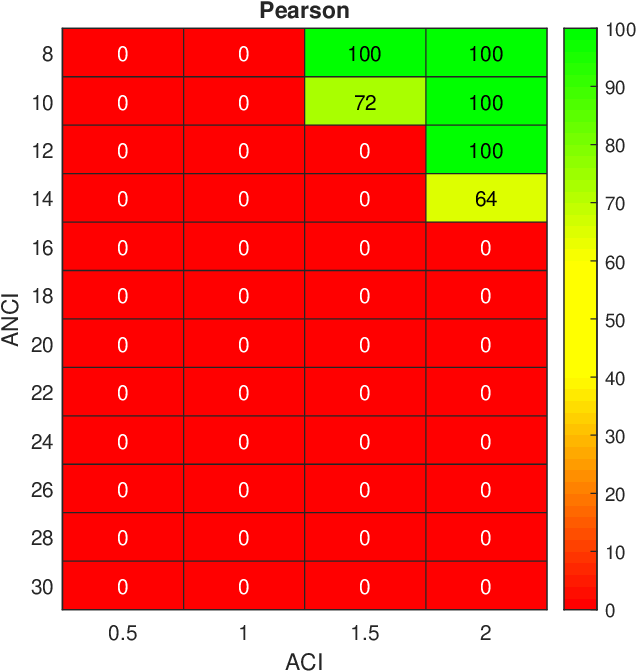}
                  \caption{}
         \label{eff_pearson}
     \end{subfigure}
        \caption{{Efficiency of the selectors: spectral kurtosis, CV, alpha, and Pearson for different values of $A_{CI}$, $A_{NCI}$ based on 50 MC simulations.}}
        \label{fig:efficiency_sel}
\end{figure}

\begin{figure}[h!]
     \centering
     \begin{subfigure}[b]{0.3\textwidth}
         \centering
         \includegraphics[width=\textwidth]{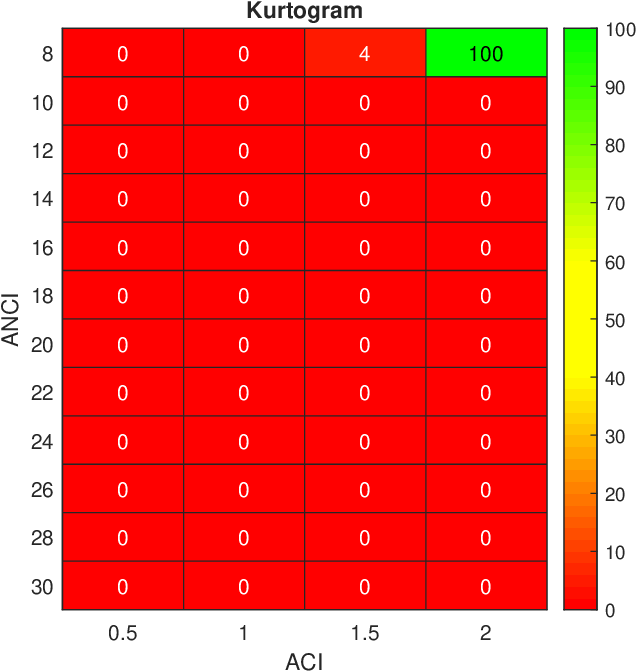}
         \caption{}
  \label{eff_kurtogram}
     \end{subfigure}
        \begin{subfigure}[b]{0.3\textwidth}
         \centering
         \includegraphics[width=\textwidth]{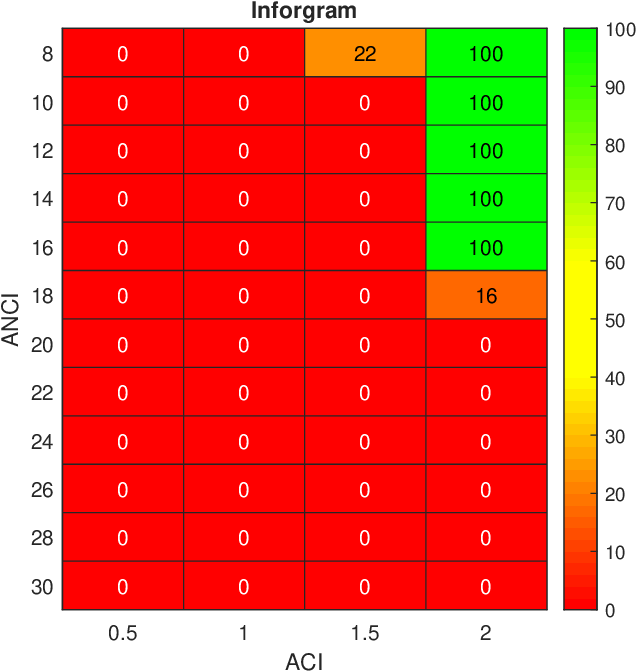}
                  \caption{}
                  \label{eff_infogram}
     \end{subfigure}
        \begin{subfigure}[b]{0.3\textwidth}
         \centering
         \includegraphics[width=\textwidth]{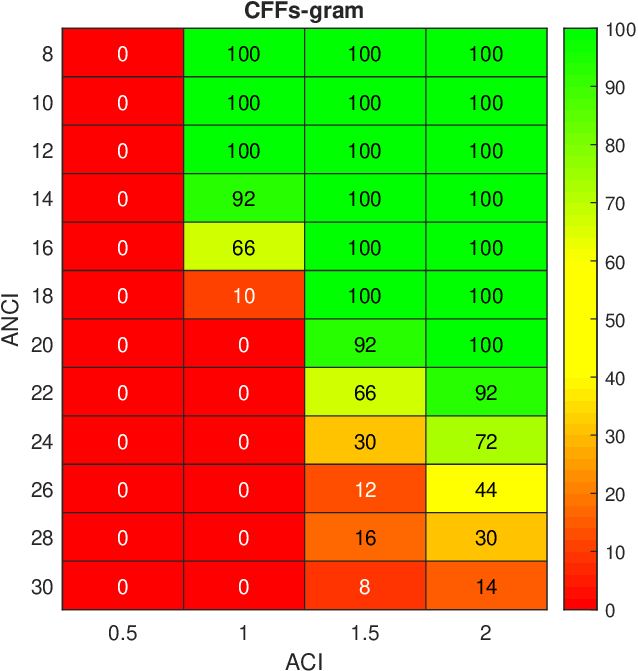}
                  \caption{}
                  \label{eff_CFFsgram}
 \end{subfigure}
        \caption{{Efficiency of the kurtogram, infogram, and CFFs-gram for different values of $A_{CI}$, $A_{NCI}$ based on 50 MC simulations.}}
        \label{fig:efficiency_grams}
\end{figure}

{Figs. \ref{fig:efficiency_sel} and \ref{fig:efficiency_grams} present the efficiency (in percent) for seven reference methods. It was obtained based on 50 Monte Carlo simulations for different values of $A_{CI}$, $A_{NCI}$.}

\end{document}